\documentclass[authoryear,3p,onecolumn,review,12pt]{elsarticle}

\usepackage{graphics}
\usepackage{natbib}
\usepackage{setspace}
\usepackage{amssymb}
\usepackage{amstext}
\usepackage{textcmds}
\usepackage{color}
\usepackage{lineno}
\usepackage{longtable}
\usepackage{tabularx}
\usepackage{array}
\usepackage[version=3]{mhchem}

\small\normalsize

\journal{Icarus}

\begin{document} 

\begin{frontmatter}

\title{Photochemistry, mixing and transport in Jupiter's stratosphere constrained by Cassini}

\author[SwRI]{V. Hue\corref{cor1}} \ead{vincent.hue$@$swri.org} 
\author[LAB1]{F. Hersant}
\author[LESIA]{T. Cavali\'e}
\author[LAB1]{M. Dobrijevic}
\author[JPL]{J. A. Sinclair}

\address[SwRI]{Southwest Research Institute, San Antonio, TX 78228, United States}
\address[LAB1]{Laboratoire d'astrophysique de Bordeaux, Univ. Bordeaux, CNRS, B18N, allée Geoffroy Saint-Hilaire, 33615 Pessac, France.}
\address[LESIA]{LESIA, Observatoire de Paris, CNRS, Universit\'e Paris 06, Universit\'e Paris-Diderot, 5 place Jules Janssen, 92195 Meudon, France}
\address[JPL]{Jet Propulsion Laboratory/California Institute of Technology, 4800 Oak Grove Dr, Pasadena, CA 91109, United States}


\cortext[cor1]{Tel: +1-210-522-5027}

\date{ Received Oct. 13, 2017; Revised Jan. 12, 2018; Accepted Feb. 13, 2018}

\begin{abstract}

In this work, we aim at constraining the diffusive and advective transport processes in Jupiter's stratosphere, using Cassini/CIRS observations published by \citet{Nixon2007, Nixon2010}. The Cassini-Huygens flyby of Jupiter on December 2000 provided the highest spatially resolved IR observations of Jupiter so far, with the CIRS instrument. The IR spectrum contains the fingerprints of several atmospheric constituents and allows probing the tropospheric and stratospheric composition. In particular, the abundances of C$_2$H$_2$ and C$_2$H$_6$, the main compounds produced by methane photochemistry, can be retrieved as a function of latitude in the pressure range at which CIRS is sensitive to. CIRS observations suggest a very different meridional distribution for these two species. This is difficult to reconcile with their photochemical histories, which are thought to be tightly coupled to the methane photolysis. While the overall abundance of C$_2$H$_2$ decreases with latitude, C$_2$H$_6$ becomes more abundant at high latitudes. In this work, a new 2D (latitude-altitude) seasonal photochemical model of Jupiter is developed. The model is used to investigate whether the addition of stratospheric transport processes, such as meridional diffusion and advection, are able to explain the latitudinal behavior of C$_2$H$_2$ and C$_2$H$_6$. We find that the C$_2$H$_2$ observations are fairly well reproduced without meridional diffusion. Adding meridional diffusion to the model provides an improved agreement with the C$_2$H$_6$ observations by flattening its meridional distribution, at the cost of a degradation of the fit to the C$_2$H$_2$ distribution. However, meridional diffusion alone cannot produce the observed increase with latitude of the C$_2$H$_6$ abundance. When adding 2D advective transport between roughly 30\,mbar and 0.01\,mbar, with upwelling winds at the equator and downwelling winds at high latitudes, we can, for the first time, reproduce the C$_2$H$_6$ abundance increase with latitude. In parallel, the fit to the C$_2$H$_2$ distribution is degraded. The strength of the advective winds needed to reproduce the C$_2$H$_6$ abundances is particularly sensitive to the value of the meridional eddy diffusion coefficient. The coupled fate of these methane photolysis by-products suggests that an additional process is missing in the model. Ion-neutral chemistry was not accounted for in this work and might be a good candidate to solve this issue.

\end{abstract}


\end{frontmatter}

\section{Introduction}
\label{s:Introduction}

The stratospheres of the giant planets are driven by the interaction between the temperature, composition and dynamics. These interactions are initiated by the incoming sunlight energy input. Photochemical reactions initiated by sunlight UV radiation make atmospheric composition more complex \citep{Moses2005b, Hue2015, Hue2016}, radiative heating by some of the photochemistry products affects atmospheric temperature, and temperature and composition gradients affect in turn atmospheric dynamics (e.g. \citealt{West1992}).

On Saturn, seasonal changes have been witnessed by Cassini, mainly on the stratospheric temperatures \citep{Fletcher2010, Sinclair2013} and more subtly on the atmospheric composition \citep{Guerlet2009,Sylvestre2015}. Jupiter, with a 3$^{\circ}$ obliquity, experiences smaller seasonal variations of incoming sunlight, when compared to the 26.7$^{\circ}$ obliquity of Saturn. In addition, the orbital modulation of the received solar radiation is also smaller for Jupiter, because of its smaller eccentricity (0.048392) compared to Saturn (0.054150). For these reasons, the seasonal effects on Jupiter are expected to be much less pronounced than for Saturn. Nevertheless, Jupiter also exhibits a seasonally-induced hemispheric asymmetry in stratospheric temperature as observed by \citet{SimonMiller2006} and \citet{Nixon2007}, although weaker than on Saturn. Meridional variability of several hydrocarbons and other species has also been observed in Jupiter's stratosphere.

Early theoretical work from \citet{Cadle1962} and \citet{Strobel1969, Strobel1973} predicted that methane photolysis leads to the formation of higher order hydrocarbons such as acetylene (C$_2$H$_2$), ethylene (C$_2$H$_4$), and ethane (C$_2$H$_6$). These predictions were confirmed by the detection of C$_2$H$_2$ and C$_2$H$_6$ by \citet{Ridgway1974} and \citet{Combes1974}. Many ground-based observations then improved the accuracy on the abundances of these molecules (see, e.g., \citet{Sada1998} for a list of observations). Observations during the Jupiter flybys of the Voyager missions suggested that C$_2$H$_2$ and C$_2$H$_6$ abundances decreased and increased, respectively, by a factor of three from the equator to the north polar region \citep{Maguire1984}. Eventually, Cassini mapped Jupiter's thermal infrared spectrum during its 2000 flyby, allowing \citet{Kunde2004} and \citet{Nixon2007} to measure the meridional variability of C$_2$H$_2$ and C$_2$H$_6$. Although C$_2$H$_6$ is a more stable molecule than C$_2$H$_2$, their meridional behavior is expected to follow the annual mean insolation, according to photochemical models without meridional transport \citep{Moses2005b, Hue2015}. However, the Cassini observations indicate that both molecules have an opposite behavior: while the C$_2$H$_2$ abundance globally decreases with latitude, the abundance of C$_2$H$_6$ increases with latitude. This trend was recently confirmed with a large set of ground based observations, showing that its feature is stable over time \citep{Melin2018}.

Since C$_2$H$_2$ and C$_2$H$_6$ are produced at high altitudes from CH$_4$ photolysis over short timescales (when compared to Jupiter's orbital period), they can be used to study stratospheric transport processes, provided that photochemistry is accounted for. Such models then allow coupling chemistry with transport processes, and have typically been used to constrain the vertical diffusion processes \citep{Gladstone1996, Moses2005a,Cavalie2008b,Cavalie2012}. Recently, the methodology based on chemical uncertainties propagation presented by \citet{Hebrard2013}, \citet{Dobrijevic2014}, and \citet{Loison2015}, has significantly improved the photochemical model predictability. Once the vertical diffusion processes and the photochemical reactions are well defined, the observed meridional variations compared to the photochemical predictions can be used to probe meridional transport. On Saturn, for instance, photochemical models have been used to indicate the regions of the stratosphere where meridional transport is likely to occur \citep{Moses2005b, Hue2015}.

Other non-hydrocarbon species in the Jovian stratosphere, like CO, CO$_2$, HCN, H$_2$O, and CS, present asymmetries in their meridional distributions as demonstrated over the past 20 years by e.g. \citet{Lellouch2002,Lellouch2006}, \citet{Moreno2003}, \citet{Griffith2004}, and \citet{Cavalie2013}. These species originate from the spectacular impacts of comet Shoemaker-Levy 9 (SL9) on Jupiter in July 1994, as they were all detected in the upper stratosphere only after the SL9 impacts \citep{Lellouch1995,Lellouch1997,Bjoraker1996,Marten1995,Bezard1997,Bezard2002}, with the exception of CO which had also been observed before \citep{Beer1975,Noll1988}. These species have presumably been formed during the comet impact by shock chemistry \citep{Zahnle1996}. Such events seem to be ubiquitous in the outer Solar System (see \citealt{Cavalie2009,Cavalie2010} for Saturn, \citealt{Cavalie2014} for Uranus, and \citealt{Lellouch2005}, \citealt{Moreno2017} for Neptune). Interestingly, tracing the evolution of the spatial distribution (both vertical and horizontal) of these rather long-lived species can provide us with an insight on the Jovian upper stratospheric dynamics. \citet{Lellouch2002}, \citet{Moreno2003}, and \citet{Griffith2004}, have estimated the magnitude of meridional diffusion from infrared and submillimeter observations of CO$_2$, CO, CS, and HCN, several years after the impacts. Over a longer period, and using Cassini observations recorded during Jupiter's flyby and earlier ground-based observations of \citet{Griffith2004}, \citet{Lellouch2006} invoked a combination of meridional diffusive and advective transport to explain the distributions of CO$_2$ and HCN. In the upper troposphere and lower stratosphere, and using observations of the SL9 debris from the Hubble Space Telescope (HST) of \citet{West1995} spread over 3.2\,years after the collision, \citet{Friedson1999} were able to estimate the magnitude of zonal mean eddy diffusion at mid-latitudes in the southern hemisphere.

In this work, we aim to constrain the diffusive and advective transport processes in Jupiter's stratosphere, using the meridional distributions of C$_2$H$_2$ and C$_2$H$_6$ retrieved by \citet{Nixon2007, Nixon2010} from Cassini/CIRS observations. To do so, we have adapted the 2D seasonal photochemical model presented in \citet{Hue2015} to Jupiter. We first present the effects of seasons on atmospheric composition using some of the results of \citet{Gladstone1996} and \citet{Moses2005a}, who constrained vertical eddy diffusion. Then, we compare the seasonal predictions with previous results obtained for Saturn by \citet{Hue2015}, without meridional transport. Using the Cassini observations of C$_2$H$_2$ and C$_2$H$_6$ of \citet{Nixon2007, Nixon2010}, and following \citet{Liang2005} and \citet{Lellouch2006}, we then present new constraints on meridional eddy diffusion with this fully coupled 2D-photochemical-diffusion model. Finally, we consider how altitude-latitude advective transport, such as stratospheric circulation cells, can affect the C$_2$H$_2$ and C$_2$H$_6$ abundances.



\section{2D seasonal photochemical modeling of Jupiter's stratosphere}
\label{s:Photochemical_modeling}

\subsection{The 2D photochemical model}
\label{ss:model}

We have adapted to Jupiter the 2D-spherical photochemical model developed for Saturn \citep{Hue2015}. The evolution of the atmospheric compounds governed by the continuity equation can be written as:

\begin{equation}
\dfrac{\partial n_i}{\partial t} = P_i - n_i L_i - \frac{1}{r^2} \dfrac{\partial \left( r^2 \Phi^{r}_i \right) }{\partial r} + \frac{1}{r \cos \theta} \dfrac{\partial \left( \cos \theta \Phi^{\theta}_i \right) }{ \partial \theta }  \label{eq:continuity}
\end{equation}

where $r$ is the radius, $\theta$ the latitude\footnote{Latitudes are all planetocentric in this paper.}, $n_i$ [cm$^{-3}$] the number density, $P_i$ [cm$^{-3}$\,s$^{-1}$] the (photo)chemical production rate, and $L_i$ [s$^{-1}$] the (photo)chemical loss rate. $\Phi^{r}_i$ and $\Phi^{\theta}_i$ [cm$^{-2}$\,s$^{-1}$] are respectively the vertical and meridional particle fluxes due to transport. The model uses a finite difference method to solve equation \ref{eq:continuity} for each compound $i$, each pressure level and each latitude. We use the DLSODES solver, part of the ODEPACK library \citep{DLSODE}. We use a uniform altitude grid with 126 levels from 1000\,mbar to 10$^{-6}$\,mbar. Consecutive altitude levels are separated by 8\,km, which is about a third of Jupiter's scale height in the middle stratosphere. A zero flux was assumed as the upper boundary condition for all species, except for atomic hydrogen H, for which a flux of -1.5\,$\times$\,10$^{9}$\,cm$^{-2}$\,s$^{-1}$ was assumed (at all latitudes), following \citet{Moses2005a}. This flux signifies the production of atomic hydrogen by photochemical processes at higher altitude, above the upper limit of the model. The photochemical model results are not very sensitive to the value of this flux \citep{Moses2005a}. At the lower boundary, CH$_4$ was set to 1.81\,$\times$\,10$^{-3}$ \citep{VonZahn1998} and the He abundance to 0.136 \citep{Niemann1998}. Finally, the H$_2$ abundance at the lower boundary was set to 0.86219, so that $y$(CH$_4$) $+$ $y$(H$_2$) $+$ $y$(He) = 1. The model boundary conditions in the latitudinal dimension assumes that $\Phi^{\theta}\,=\,0$, meaning that there are no particles coming in and out of the lower latitudinal boundary of the first latitudinal cell. The same thing stands for the last latitudinal cell, i.e. there are no particles coming in and out of the upper latitudinal boundary of the last latitudinal cell.

Post-SL9 observations of the spreading debris (see, e.g., \citealt{West1996}, \citealt{Banfield1996}, \citealt{Friedson1999}) have suggested that longitudinal transport occurs faster than the latitudinal one in Jupiter's atmosphere at the pressures levels (1-400\,mbar) where the debris were deposited. We assume here that the longitudinal transport at the submillibar pressure levels operates as fast as at higher pressure levels.

The vertical and meridional fluxes of chemical species are defined as follow:

\begin{align}
\Phi^{r}_i =& - D_i n_i \left( \frac{1}{y_i} \dfrac{\partial y_i}{\partial r} + \frac{1}{H_i} - \frac{1}{H} \right) - K_{zz} n_i \left( \frac{1}{y_i} \dfrac{\partial y_i}{\partial r} \right)  + v_i^r n_i \label{eq:flux_z} \\
\Phi^{\theta}_i =& - (D_i + K_{yy}) \, n \left( \dfrac{\partial y_i}{r \partial \theta} \right) + v_i^{\theta} n_i \label{eq:flux_theta}
\end{align}

where $y_i$ is the abundance of species $i$, defined as the ratio between the number density of $i$ over the total number density. $H_i$ and $H$ [cm] are respectively the species $i$ and the mean density scale heights, and $D_i$ [cm$^{2}$\,s$^{-1}$] the molecular diffusion coefficient. $K_{zz}$ and $K_{yy}$ [cm$^{2}$\,s$^{-1}$] are respectively the vertical and meridional eddy diffusion coefficient, while $v_i^r$ and $v_i^{\theta}$ [cm\,s$^{-1}$] are the advective vertical and meridional winds, respectively. Equation \ref{eq:flux_z} shows that the molecular diffusion contains a purely diffusive part ($\propto \partial y_i / \partial r$) and a purely advective part ($\propto 1/H_i - 1/H$). Due to the unstable nature of the numerical treatment of advection, a first-order upwind scheme was used to treat the advective part of the molecular diffusion \citep{Godunov1959}. A second order upwind scheme, coupled with the OSPRE flux limiter, was used to treat the advective transport (see the review of \citealt{Waterson2007}).


In what follows, we describe the input parameters used in the photochemical model.

\subsection{Temperature profile}
\label{sss:Temperature_Prof}

The amount of received sunlight drives the stratospheric temperature. Because of Jupiter's small obliquity, seasonal effects on the stratospheric temperature are less important than in Saturn. 
\begin{itemize}
  \item In the upper troposphere, around 200\,mbar, the north/south meridional gradient as seen from Cassini was negligible on Jupiter \citep{Nixon2007} while it was about 10\,K on Saturn at that same pressure level in 2005 \citep{Fletcher2007b}.
  \item The stratospheric temperature meridional variations on Jupiter are much less pronounced than the ones observed on Saturn. Cassini/CIRS has measured a 7-8\,K north/south temperature gradient at 5\,mbar \citep{Nixon2007} during the Jupiter flyby, while this gradient was about 40\,K at 2.1\,mbar on Saturn in 2005 \citep{Sinclair2013}. The variations between the temperatures measured by Voyager/IRIS and Cassini/CIRS, measured $\mathtt{\sim}$ 21.8\,years apart (1.76 Jovian years) are of $\sim$10\,K in the lower stratosphere (at 10\,mbar) and remain below 15\,K in the upper stratosphere (from 1\,mbar to 0.01\,mbar) \citep{SimonMiller2006, Nixon2010}. 
\end{itemize}
Our simulations show that such hemispheric temperature differences affect the 5-mbar abundances of C$_2$H$_6$ and C$_2$H$_2$ by only 5 and 20\%, respectively, which is much less than the equator-to-pole variations seen by \citet{Nixon2007}.

Therefore, we have adopted uniformly at all latitudes the thermal profile of \citet{Fouchet2000}, which is based on the Galileo entry probe measurements \citep{Seiff1998} with modifications between 30\, and 10$^{-3}$\,mbar to fit ISO/SWS observations. They have smoothed the vertical oscillations of the temperature caused by Jupiter's quasi-quadrennial oscillation (see \citealt{Greathouse2016} for pluri-annual high resolution observations of these oscillations). Infrared observations indicate that Jupiter's thermal field shows longitudinal inhomogeneities. These inhomogeneities are likely caused by vertically propagating waves (e.g. \citealt{Leovy1991, Orton1991}) and are obviously not accounted for in our latitude-altitude model.

\subsection{Vertical eddy diffusion coefficient profile}
\label{sss:Temperature_Prof}

\begin{figure}[!h]
\centering
{\includegraphics[width=0.6\columnwidth]{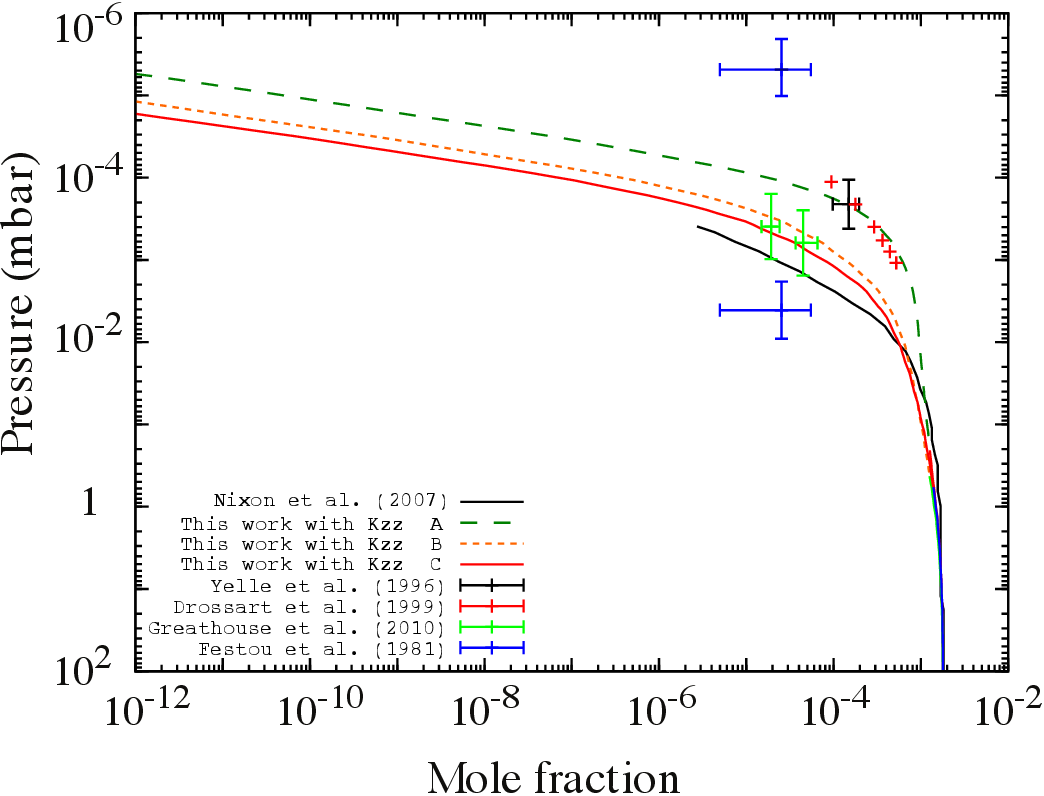}}
\caption{CH$_4$ vertical profiles predicted with various eddy diffusion coefficients ($K_{zz}$). The black line denotes the a priori CH$_4$ vertical profile \citet{Nixon2007} used in their retrieval. The red solid line denotes our model results using the model C $K_{zz}$ profile of \citet{Moses2005a}. We use this profile in this work. CH$_4$ vertical profiles obtained using Moses et al.'s models A and B $K_{zz}$ profiles are also presented, in dashed green and solid orange lines respectively. The various CH$_4$ observations come from \citet{Festou1981, Yelle1996, Drossart1999, Greathouse2010}.}
\label{fig:Kzz_Jup}
\end{figure}

The vertical eddy diffusion coefficient $K_{zz}$ used here corresponds to the one from model C of \citet{Moses2005a}. The CH$_4$ vertical profiles predicted by the photochemical model using the $K_{zz}$ profiles of \citet{Moses2005a} are presented in Fig. \ref{fig:Kzz_Jup}. The different observational constraints suggest a rather broad range of pressure levels for the position of the methane homopause. These levels were obtained using different techniques, and retrieved for different latitudes and dates. There is no clear evidence yet for a possible latitudinal and/or temporal variation in the methane homopause position, and we use at all latitudes the $K_{zz}$ from model C of \citet{Moses2005a} which allows a satisfactory reproduction of the CH$_4$ observations recorded by New Horizons/UVS \citep{Greathouse2010}.

\subsection{Jupiter's intrinsic and orbital characteristics}
\label{sss:Jup_Characteristics}

\begin{figure}[!h]
\centering
{
\includegraphics[width=0.9\columnwidth]{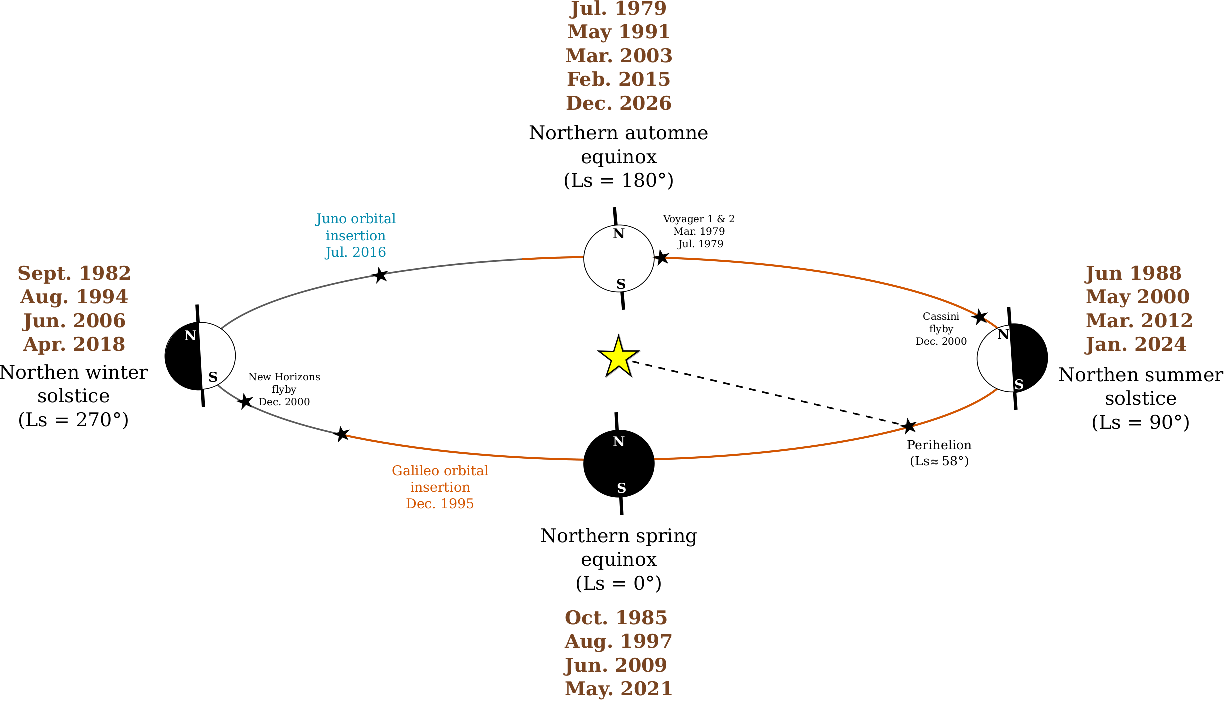} \\
\includegraphics[width=0.8\columnwidth]{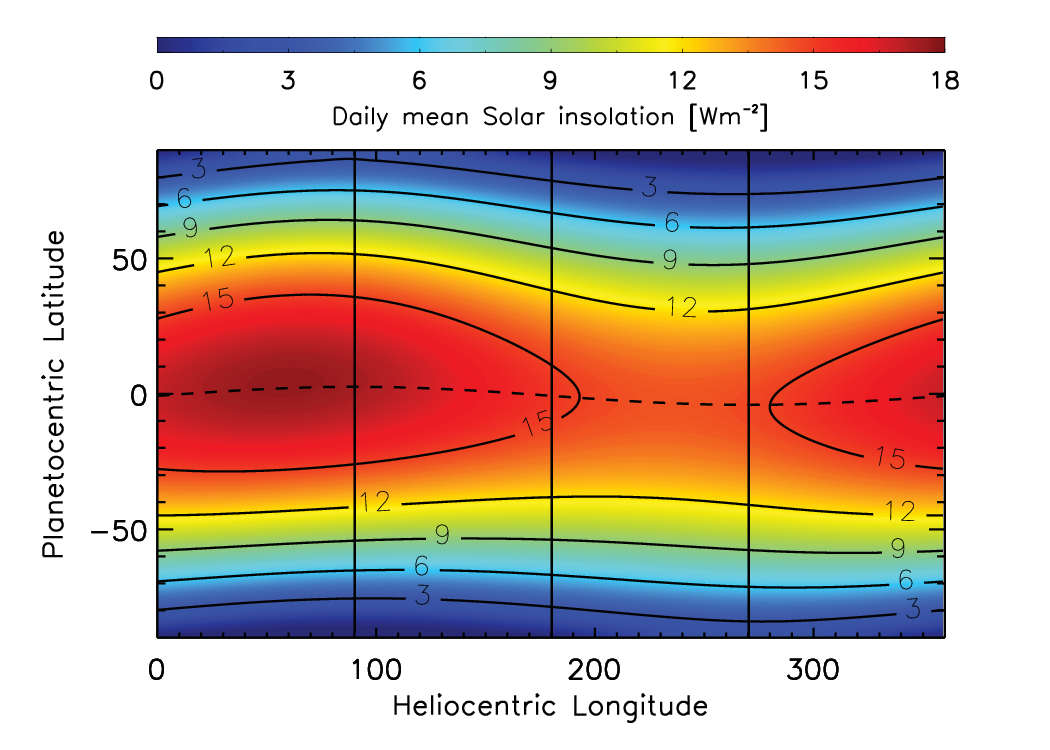}
}
\caption{Top panel: overview of Jupiter's orbit. The space missions that have visited the planet are indicated. 
Bottom panel: Daily mean insolation (in W$\cdot$m$^{-2}$) as a function of planetocentric latitude and heliocentric longitude ($L_s$, i.e., seasons) received by a horizontal unit surface in Jupiter's atmosphere. Jupiter's perihelion occurs at $L_s$ $\approx$ 58$^{\circ}$, based on J2000.}
\label{fig:Orbital_Jup}
\end{figure}

Jupiter orbits the sun in 11.86\,years. Its orbit is such that Jupiter's perihelion occurs during its northern spring season, hence making the northern spring and summer shorter than their southern counterparts. The upper panel of Fig. \ref{fig:Orbital_Jup} presents an overview of Jupiter's seasons, and shows the different space missions that visited this planet. The evolution of the daily mean insolation as a function of planetocentric latitude and heliocentric longitude is presented at the bottom of Fig. \ref{fig:Orbital_Jup}. It reflects the combined effect of the planet's obliquity and eccentricity.

\subsection{Chemical network}
\label{sss:Chemical_network}

Although the physical conditions in the giant planets are different (heliocentric distance, stratospheric temperature, vertical mixing, boundary conditions), the CH$_4$ photolysis initiates and controls their stratospheric hydrocarbon chemistry (e.g., \citealt{Gladstone1996}, \citealt{Moses2005a}, and references therein). The seasonal variations of chemical composition for a given latitude are then likely to depend on the planet obliquity.

\begin{figure}[!h]
\centering
{
\includegraphics[width=0.45\columnwidth]{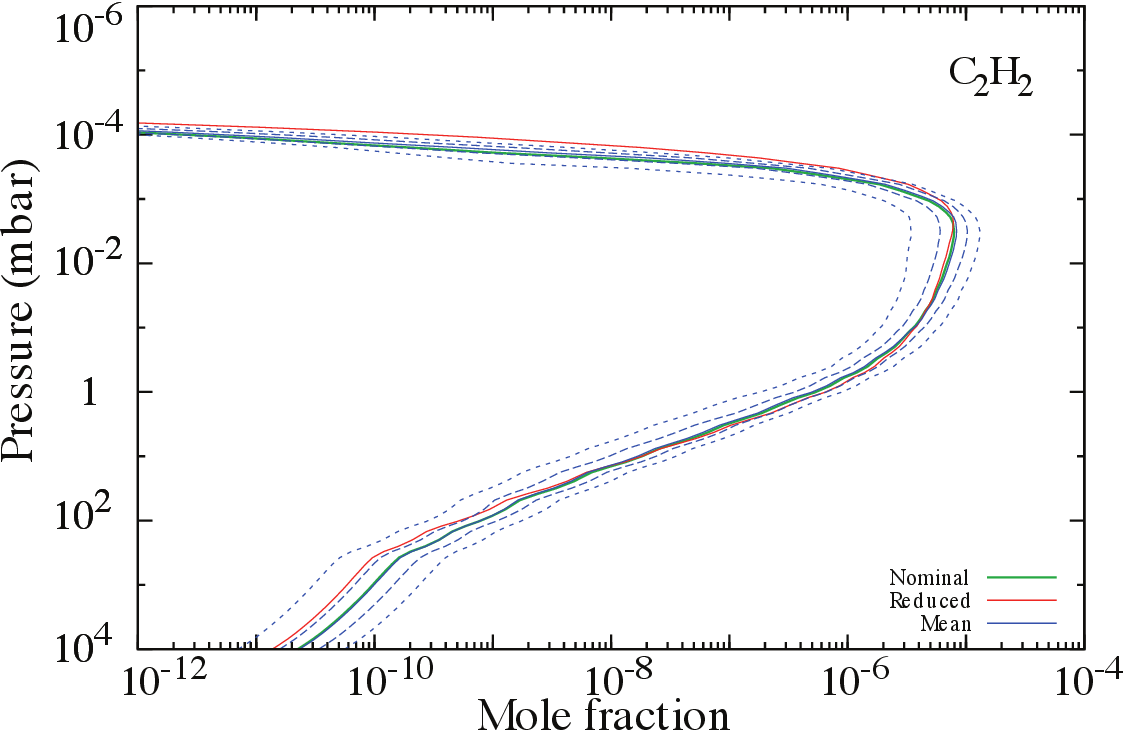}
\includegraphics[width=0.45\columnwidth]{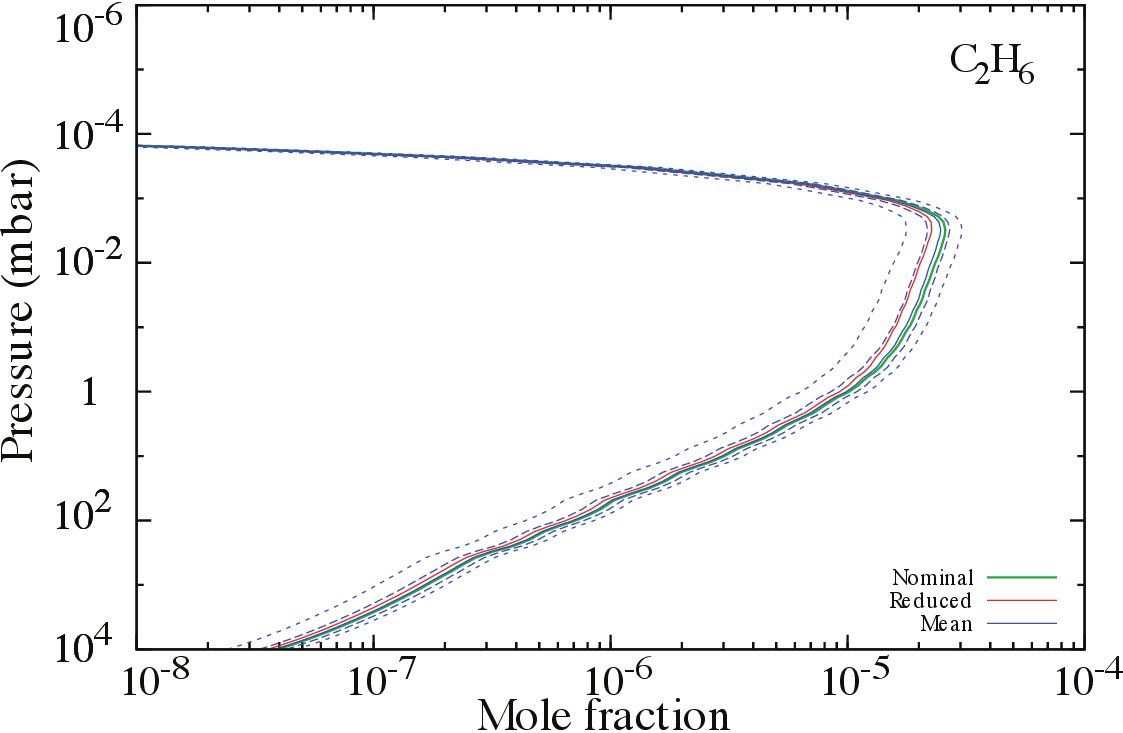}
}
\caption{Vertical profiles of C$_2$H$_2$ (left panel) and C$_2$H$_6$ (right panel) in Jupiter's stratosphere. The red lines correspond to the profiles computed with the reduced chemical network of \citet{Hue2015}. This network was reduced in Saturn's physical conditions and run here under Jupiter's condition. The green lines stand for the profiles computed with the nominal network of \citet{Loison2015}. 500 runs of uncertainty propagation have been performed on the nominal network's reaction rates. The 1st and 19th 20-quantiles of the distribution, and the 5th and 15th 20-quantiles of the distribution (respectively), are plotted in blue short-dashed lines, and blue long-dashed lines (respectively). The mean vertical profile is displayed with a solid blue line.
}
\label{fig:Reduction}
\end{figure}

Our chemical network is based on the network developed by \citet{Loison2015}. Following earlier work of \citet{Dobrijevic1998}, \citet{Dobrijevic2003,Dobrijevic2008,Dobrijevic2010} and \citet{Hebrard2006,Hebrard2007,Hebrard2009}, they have reviewed and improved the C, N, and O, neutral chemistry to study Titan's atmosphere. Such typical 1D photochemical models use chemical networks that include as many chemical compounds and reactions as possible. The size of these networks (hundreds compounds coupled in about a thousand reactions) makes 1D photochemical models then hardly extendable to 2 or 3D, because one has to solve equation \ref{eq:continuity} for each compound, each pressure level, and each latitude. This has led \citet{Dobrijevic2011} to develop an objective methodology to reproduce the chemical processes for a subset of compounds of interest (compounds that are usually observed) with a limited number of reactions. This subset is extracted from a more complex chemical network by running a 1D photochemical model, applying uncertainty propagations on the chemical reaction rates, and a global sensitivity analysis. \citet{Hue2015} have reduced the chemical network of \citet{Loison2015} in conditions relevant to Saturn's stratosphere. Adapting the 2D-photochemical model of \citet{Hue2015} to Jupiter raises then the following question: is the reduced network developed in Saturn's physical conditions also valid in Jupiter's conditions? To answer this question, we have run an uncertainty propagation study on the chemical reaction rates of \citet{Loison2015} under Jupiter stratospheric conditions, and we have compared the abundance vertical profiles of the species of interest, as computed with the nominal and reduced networks. The results are presented in Fig. \ref{fig:Reduction}.

The C$_2$H$_2$ vertical profile predicted with the reduced network is always within the 1st and the 19th 20-quantiles of the nominal network distribution, except at pressures lower than 5\,$\times$\,10$^{-4}$\,mbar, i.e. above the methane homopause. In the stratosphere, down to 100\,mbar, the C$_2$H$_2$ vertical profile is between the 5th and the 15th 20-quantiles of the distribution. At pressures larger than 100\,mbar, the C$_2$H$_2$ vertical profile is between the 1st and the 5th of the distribution. The C$_2$H$_6$ photochemistry is simpler than the C$_2$H$_2$ one. This results in the C$_2$H$_6$ profile to always remain within the 5th and the 15th 20-quantiles of the distribution. The profiles computed with the reduced network of \citet{Hue2015} are thus accurate enough when this network is used under Jupiter conditions and we proceed with it in this study.

\section{Seasonal model results without meridional diffusion and transport}
\label{s:seasonal_without_diffusion}

We present here the results obtained with the photochemical model without meridional diffusion ($K_{yy}$ $=$ 0) and without 2D advective transport ($v^r$ and $v^{\theta}$ $=$ 0). Since we use a reduced chemical network whose purpose is to accurately reproduce the chemical abundances of the observed C$_2$H$_x$ compounds, we will only discuss these species hereafter.

\begin{figure}[!h]
\centering
{\includegraphics[width=0.6\columnwidth]{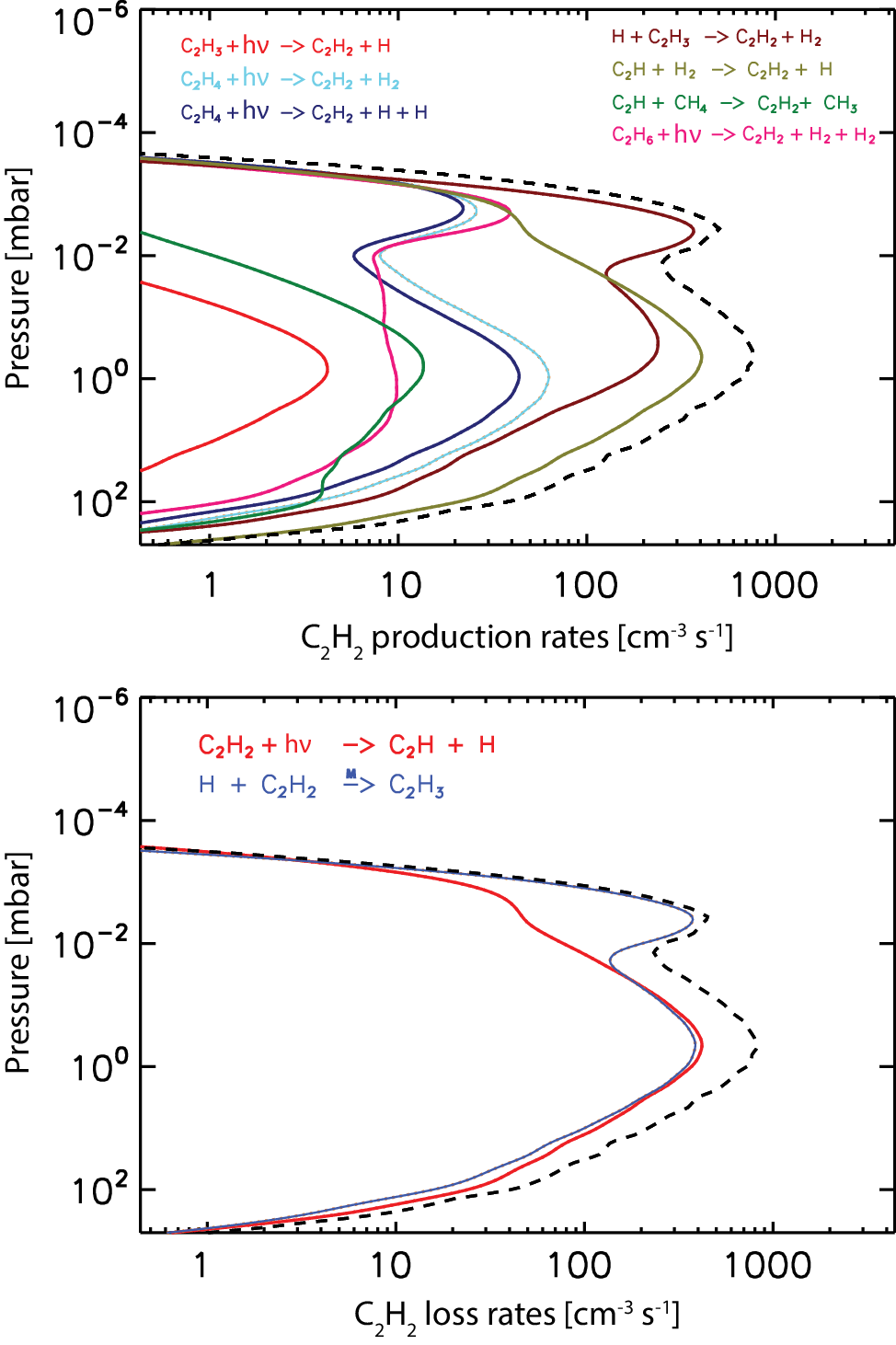}
}
\caption{Production rates (top panel) and loss rates (bottom panel) of the main reactions leading to the production and destruction of C$_2$H$_2$. These rates are given at the equator and during Jupiter's northern spring equinox ($L_s$ = 0$^{\circ}$). The black dashed line represents the total production and loss rates.}
\label{fig:Prod_Perte_C2H2_jup}
\end{figure}

The main reactions leading to the production and destruction of C$_2$H$_2$ are presented in Fig. \ref{fig:Prod_Perte_C2H2_jup}. They can be compared to previous 1D photochemical model of \citet{Moses2005a}. Moses et al. predicts that C$_2$H$_2$ is mainly produced from the photolysis of C$_2$H$_4$ and C$_2$H$_6$ throughout the Jovian atmosphere. Our model predicts that C$_2$H$_2$ is mainly produced from the H + C$_2$H$_3$ $\rightarrow$ C$_2$H$_2$ + H$_2$ reaction in the higher stratosphere, down to 2\,$\times$\,10$^{-2}$\,mbar, while it is mainly produced from the C$_2$H  + H$_2$ $\rightarrow$ C$_2$H$_2$ + H reaction at higher pressures. The main reactions controlling the C$_2$H$_2$ production in this model are identical to the ones in the Saturn model of \citet{Hue2015}. Our model is consistent with \citet{Moses2005a} regarding the main reactions causing the destruction of C$_2$H$_2$, i.e. mainly by photolysis and three-body reactions with H atoms.

\begin{figure}[!h]
\centering
{\includegraphics[width=0.6\columnwidth]{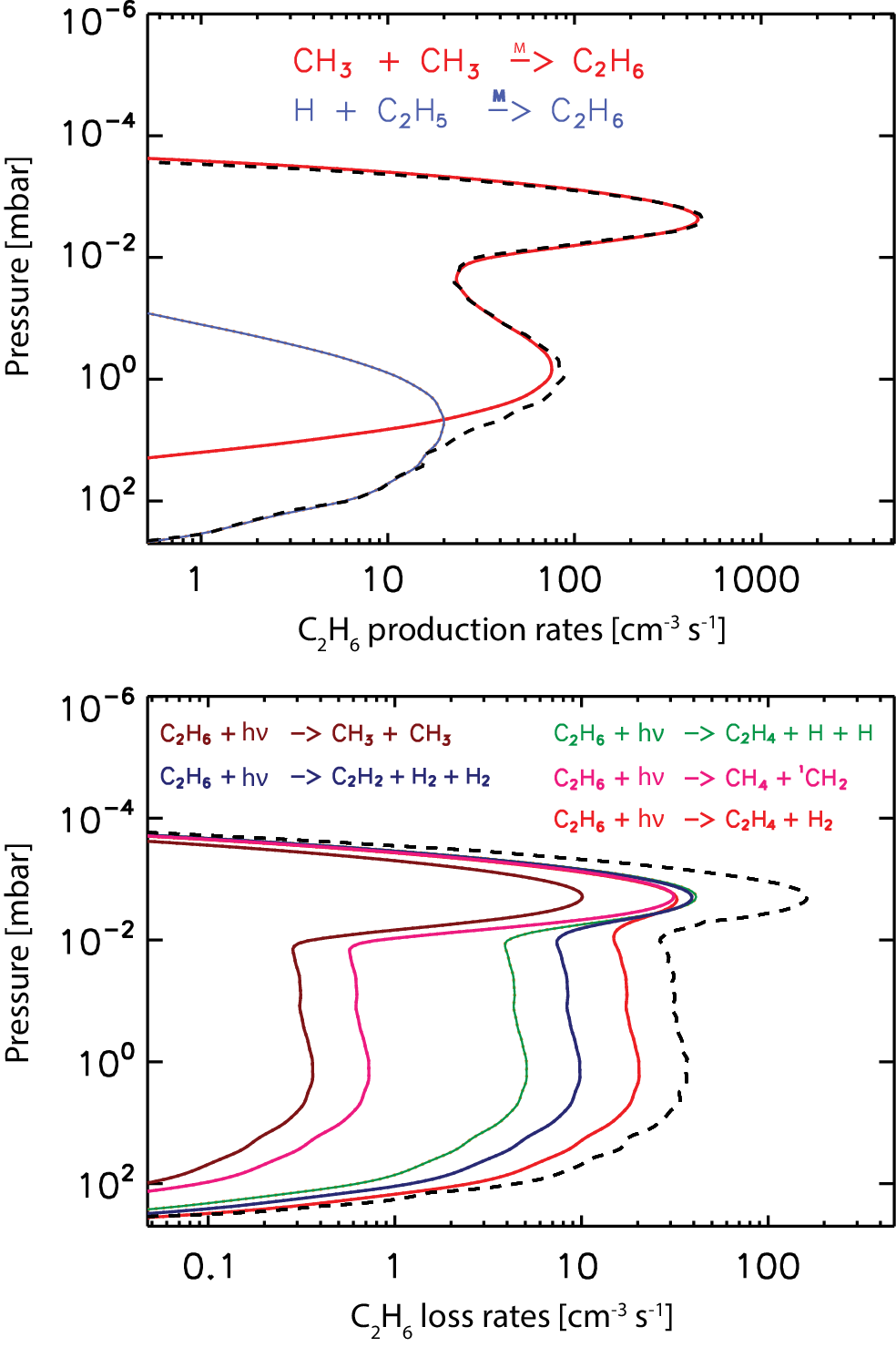}
}
\caption{Production rates (top panel) and loss rates (bottom panel) of the main reactions leading to the production and destruction of C$_2$H$_6$. These rates are given at the equator and during Jupiter's northern spring equinox ($L_s$ = 0$^{\circ}$). The black dashed line represents the total production and loss rates.}
\label{fig:Prod_Perte_C2H6_jup}
\end{figure}

From the upper stratosphere down to 5\,mbar, C$_2$H$_6$ is mainly produced from the methyl-methyl (CH$_3$) reaction, as in \citet{Moses2005a}. At higher pressure, C$_2$H$_6$ is produced from the three-body reaction in which an H atom is added to C$_2$H$_5$ (see Fig. \ref{fig:Prod_Perte_C2H6_jup}). Consistently with \citet{Moses2005a}, C$_2$H$_6$ is mainly destroyed through photolysis. As for C$_2$H$_6$, the main production and destruction channels for C$_2$H$_2$ are similar to the ones in the Saturn model of \citet{Hue2015}.

\begin{figure}[!h]
\centering
{\includegraphics[width=0.9\columnwidth]{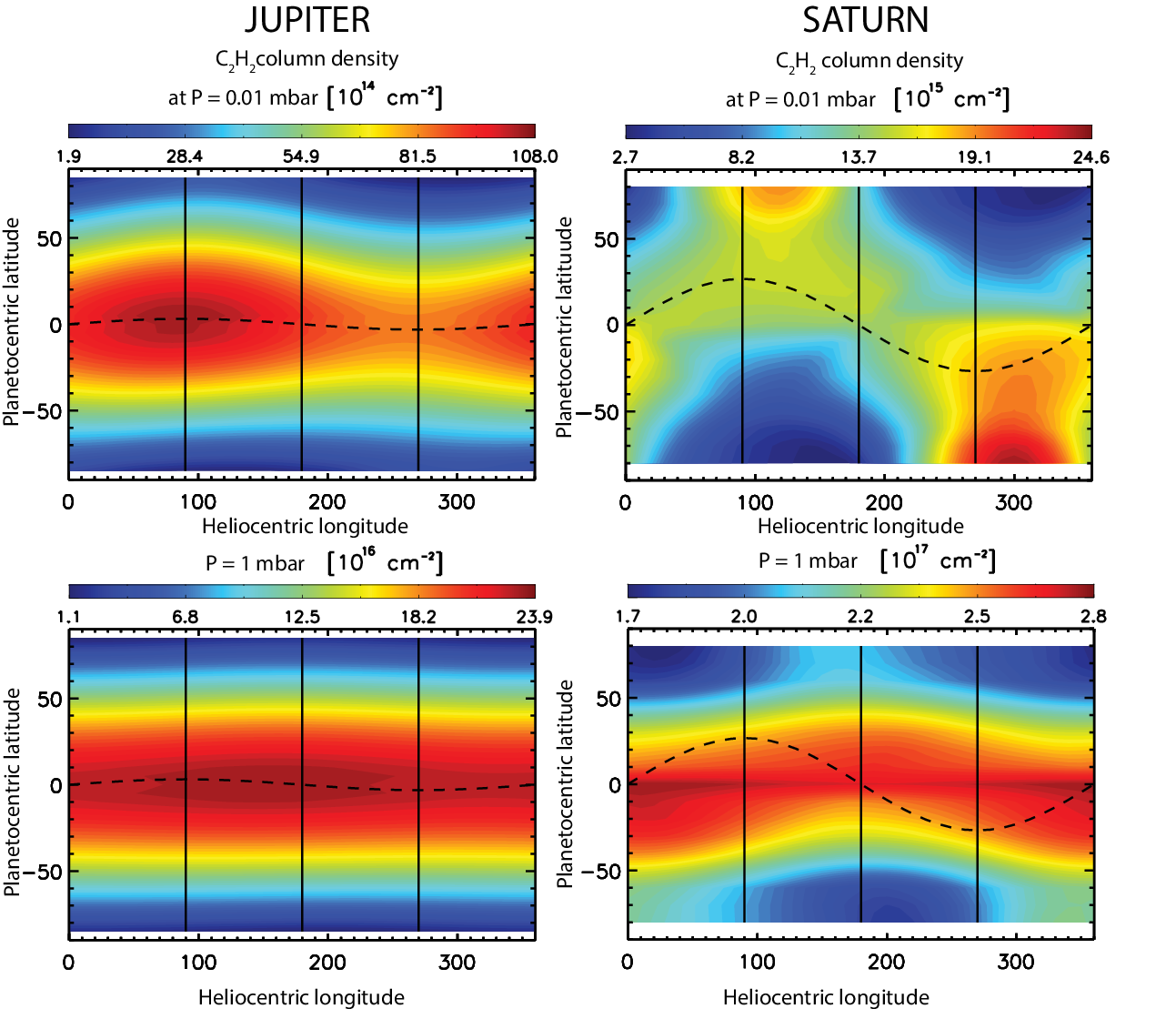}
}
\caption{Seasonal evolution of the C$_2$H$_2$ column densities on Jupiter (left panels) and Saturn (right panels), at 10$^{-2}$\,mbar (top) and 1\,mbar (bottom). The black vertical lines denote the position of the solstices and equinoxes. The dashed lines represent the evolution of the subsolar latitude on Jupiter and Saturn.}
\label{fig:Seasons_Jup}
\end{figure}

Due to Jupiter's low eccentricity, the seasonal evolution of the stratospheric composition in the present model shows less variations than in Saturn. Fig. \ref{fig:Seasons_Jup} compares the seasonal evolutions of Saturn's and Jupiter's C$_2$H$_2$ column densities at 10$^{-2}$ and 1\,mbar, without meridional diffusive and advective transport. In Saturn, the seasonal evolution of the C$_2$H$_2$ column density at 10$^{-2}$\,mbar is mainly driven by its obliquity and shows a seasonal modulation caused by its eccentricity. Note that Saturn's perihelion occurs at L$_S$\,=\,280$^{\circ}$. At higher pressure levels, when the chemical timescales progressively decrease with increasing pressure, the chemical abundances become progressively controlled by the vertical diffusion. The seasonal evolution of the column density at 1\,mbar therefore follows the yearly average received solar flux rather that the daily averaged flux, consistent with \citet{Moses2005b}. In Jupiter, the seasonal evolution of the C$_2$H$_2$ column is mostly controlled by Jupiter's eccentricity and it reaches its seasonal maximum value at L$_S$\,=\,80$^{\circ}$ at 10$^{-2}$\,mbar, i.e. after its perihelion (L$_S$\,=\,58$^{\circ}$). This maximum is shifted to L$_S$\,=\,150$^{\circ}$ at 10\,mbar because of the increase in the chemical timescales. The same reasoning stands for C$_2$H$_6$, although less pronounced because it has a longer chemical lifetime than C$_2$H$_2$.


Despite Jupiter's and Saturn's similarities in terms of chemical composition, the photochemical predictions for these two planets show striking differences due to their intrinsic (e.g. ring system, obliquity) and orbital characteristics. When compared to observational data, such predictions can help us to constrain the model free parameters. In this new type of 2D model, these free parameters are the meridional diffusion coefficient ($K_{yy}$) and the 2D advective transport ($v^{r}$ and $v^{\theta}$ = 0).

\section{Constraining diffusive and advective transport from Cassini flyby observations}
\label{s:Mixing_Transport}

The Cassini flyby of Jupiter which occurred on December 2000 has allowed us to remotely probe its atmosphere in the infrared spectral range with the CIRS instrument \citep{Flasar2004} with an unprecedented spatial resolution at the time. Several thousands of spectra taken at the highest CIRS spectral resolution ($\Delta \nu$ = 0.48\,cm$^{-1}$) were recorded between the 600-1500 cm$^{-1}$ range, which contains the fingerprints of Jupiter's main upper-tropospheric and stratospheric hydrocarbons.

\citet{Nixon2007} retrieved the abundances of C$_2$H$_2$ between 0.1 an 200\,mbar, and C$_2$H$_6$ between 5 and 200\,mbar. They have showed that the C$_2$H$_2$ and C$_2$H$_6$ meridional distributions are anti-correlated in the 5\,mbar region. While the C$_2$H$_2$ abundance globally decreases with latitude, the C$_2$H$_6$ abundance increases with latitude, hence confirming the earlier findings of \cite{Maguire1984} using Voyager/IRIS observations. This result is at odds with photochemical model predictions. The production of C$_2$H$_2$ and C$_2$H$_6$ originates from the methane photolysis. Although C$_2$H$_6$ is a more stable molecule than C$_2$H$_2$, their meridional behavior was expected to follow the annual mean insolation level, according to 1D or 2D photochemical model without meridional transport \citep{Moses2005b, Hue2015}. In a more recent paper, \cite{Nixon2010} re-analyzed the Cassini/CIRS observations along with the Voyager/IRIS observations, using an updated gas line spectral database. The differences with respect to their previous work \citep{Nixon2007} reaches up to 40\% for C$_2$H$_6$ at 5\,mbar and 1\% for C$_2$H$_2$, but confirms the opposite meridional trends of these two species. 

In this section, we aim at constraining the diffusive and advective transport processes in Jupiter's stratosphere by attempting to reproduce the CIRS observations. Since only few constraints exist, we choose to separate this study into two parts. We first consider horizontal diffusive mixing only. Then, we add advective transport to the model.

\subsection{Horizontal eddy diffusion mixing}
\label{ss:Mixing}

\subsubsection{Previous estimates and tested values}
\label{sss:previous_work}

In the lower stratosphere, between 200\,mbar and 1\,mbar, UV observations acquired with HST/WFPC-2 demonstrated that the migration of the dust generated by the SL9 impacts could not be properly modelled by the circulation pattern predicted by \citet{West1992}. \citet{Friedson1999} indeed showed that the dust migration observed more than 3 years after impact was better modelled with a purely diffusive mechanism.

Higher in the stratosphere, \citet{Lellouch2002} developed a 1D-meridional diffusion model to interpret their ISO and SWAS observations. They coupled this diffusion model to a simplified oxygen chemical scheme based on the more complex oxygen chemistry of \citet{Moses2000a,Moses2000b}. At the pressure level of 0.5\,mbar, they concluded that the post-SL9 observations performed three years after the impact could be reproduced assuming a meridional eddy diffusion coefficient of $K_{yy}$(0.5\,mbar) = 2\,$\times$\,10$^{11}$ \,cm$^{2}$\,s$^{-1}$, constant with latitude.

Several ground-based campaigns have led to other estimates of this coefficient in the upper stratosphere. \citet{Moreno2003} used millimeter and submillimeter observations with the IRAM-30m and JCMT-15m telescopes to perform a long-term monitoring of the SL9-related chemical species. Their observations ranged from 10 months up to about 4 years after the impact. They used a 2D-diffusion (latitude-longitude) model to reproduce the observations, while neglecting vertical transport. At a pressure level of 0.2\,mbar, the meridional eddy diffusion coefficient they used to reproduce the spatial spreading observed over time is $K_{yy}$(0.5\,mbar) = 2.5\,$\times$\,10$^{11}$ \,cm$^{2}$\,s$^{-1}$ in line with the \citet{Lellouch2002} value.

Infrared observations taken with IRTF/Irshell up to five years after the impact suggested a slightly different picture \citep{Griffith2004}. The observations targeted the HCN IR-emission lines, as a follow-up to the observational campaign performed shortly after the impact by \citet{Bezard1997}. In order to reproduce the meridional spreading seen in the HCN column density, \citet{Griffith2004} used a 1D diffusion model in the meridional direction, and had to invoke a meridional diffusion coefficient that varies with latitude, and with a typical magnitude of $K_{yy}$ $\approx$ 10$^{10}$\, - \,10$^{11}$\,cm$^{2}$s$^{-1}$.

\citet{Liang2005} have adapted the ``KINETICS'' 1D photochemical code (e.g. \citealt{Allen1981}, \citealt{Yung1984}, \citealt{Gladstone1996}) to run it in a quasi-two-dimensional mode. Such model allowed them to couple the 1D atmospheric columns through the addition of a meridional eddy diffusion coefficient $K_{yy}$. In order to reproduce the early Jupiter flyby Cassini data of \citet{Kunde2004}, \citet{Liang2005} found that a pressure threshold was needed, for which the efficiency of the meridional mixing drops by one order of magnitude for the lower pressures. The threshold they found was around 5-10\,mbar, which corresponds roughly to the pressure level where the meridional component of the residual circulation of \citet{Friedson1999}'s model reverses. However, they could not reconcile their low values for $K_{yy}$ in the upper stratosphere with those inferred by \citet{Lellouch2002} and \citet{Moreno2003}.

During its Jupiter flyby, Cassini provided a new insight on the meridional distribution of HCN and CO$_2$ with the CIRS instrument. The meridional distribution of these two molecules showed striking differences, CO$_2$ being concentrated around the south pole while HCN peaked at 45$^{\circ}$S and sharply decreased towards the south pole. That led \citet{Lellouch2006} to conclude that these distributions were caused by a combination of meridional diffusive and advective transport. This will be further discussed in section \ref{ss:Transport}.

\renewcommand{\arraystretch}{1.2}
\begin{table}[h!]
\footnotesize
\begin{center}
\begin{tabular}{l l l l l}
\hline
Observed & Instrument & Pressure       & Diffusion coefficient    & References  \\
 media &  & level  & retrieved   &   \\
\hline
\hline
 Dust migration & HST/WFPC-2 & 200-1\,mbar  & $K_{yy}^{max}(\theta)$ $\approx$ 10$^{11}$\,cm$^{2}$s$^{-1}$  & \citet{Friedson1999}  \\
 & &  & $K_{yy}^{min}(\theta)$ $\approx$ 10$^{10}$\,cm$^{2}$s$^{-1}$ &   \\
 &   &   &   &   \\
H$_2$O, CO$_2$ & ISO, SWAS & 0.5\,mbar & $K_{yy}$ = 2\,$\times$ 10$^{11}$\,cm$^{2}$s$^{-1}$ & \citet{Lellouch2002}  \\
and CO & and IRAM-30m  &  &  &   \\
 &   &   &   &   \\
CO, CS & IRAM-30m & 0.2\,mbar & $K_{yy}$ = 2.5 $\times$ 10$^{11}$\,cm$^{2}$s$^{-1}$ & \citet{Moreno2003}  \\
and HCN & and JCMT-15m  &   &   &   \\
 &   &   &   &   \\
HCN & IRTF-Irshell & 0.2-0.5\,mbar & $K_{yy}^{max}(\theta)$  $\approx$ 3 $\times$ 10$^{11}$\,cm$^{2}$s$^{-1}$ & \citet{Griffith2004}  \\
 &  &  &  $K_{yy}^{min}(\theta)$ $\approx$ 3 $\times$ 10$^{10}$\,cm$^{2}$s$^{-1}$  &   \\
 &   &   &   &   \\
C$_2$H$_2$ and C$_2$H$_6$ & Cassini/CIRS  & 5\,mbar & $K_{yy}$ = 2 $\times$ 10$^{9}$\,cm$^{2}$s$^{-1}$ & \citet{Liang2005}  \\
 &   &   &   &   \\
HCN, CO$_2$ & Cassini/CIRS & 0.2-0.5\,mbar & $K_{yy}^{max}(\theta)$ $\approx$ 3 $\times$ 10$^{11}$\,cm$^{2}$s$^{-1}$ & \citet{Lellouch2006}  \\
 &   &   & $K_{yy}^{min}(\theta)$ $\approx$ 5 $\times$ 10$^{9}$\,cm$^{2}$s$^{-1}$  &   \\
 &   &   & + $v^{\theta}$ (see section \ref{ss:Transport})  &   \\
\hline
\end{tabular}
\end{center}
\caption{Observational constraints on the meridional mixing processes in Jupiter's stratosphere.}
\label{fig:tab_diffusion_SL9}
\normalsize
\end{table}

Table \ref{fig:tab_diffusion_SL9} summarizes the estimates of the meridional eddy diffusion coefficient $K_{yy}$ presented above. The set of $K_{yy}$ coefficients we will use in this paper, and which are based on the aforementioned studies, is the following:

\renewcommand{\arraystretch}{1.2}
\begin{center}
\begin{tabular}{l l l}
$\bullet$  & $K_{yy}^{(1)}$($p$) & = $10^{5}$ $\times$ $K_{zz}$($p$) \\
$\bullet$  & $K_{yy}^{(2)}$($p$) & = $10^{6}$ $\times$ $K_{zz}$($p$) \\
$\bullet$  & $K_{yy}^{\mathrm{(Liang)}}$($p$) & = 2 $\times$ $10^{10}$ cm$^{2}$\,s$^{-1}$  for $p$ $\geq$ 5\,mbar \\
 &  & = 2 $\times$ $10^{9}$ cm$^{2}$\,s$^{-1}$  for $p$ \textless \, 5\,mbar \\
$\bullet$  & $K_{yy}^{\mathrm{(Lellouch 1)}}$($p$) & = 2 $\times$ $10^{11}$ cm$^{2}$\,s$^{-1}$ at $p$ = 0.3\,mbar \\
  &  & $\propto$  $K_{zz}$ for all other pressure levels \\
$\bullet$  & $K_{yy}^{\mathrm{(Lellouch 2)}}$ & = 2 $\times$ $10^{11}$ cm$^{2}$\,s$^{-1}$ \\
\end{tabular}
\label{tab:tab_Kyy}
\end{center}

\begin{figure}[!h]
\centering
{\includegraphics[width=0.65\columnwidth]{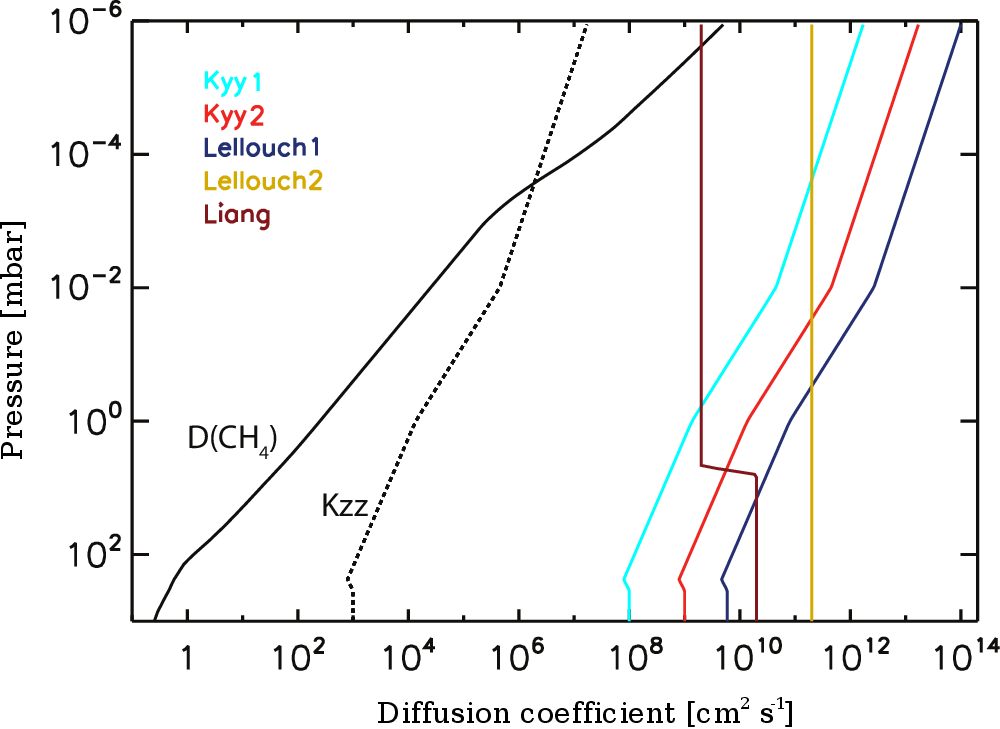}
}
\caption{Meridional diffusion coefficients $K_{yy}$ as a function of pressure, used in this study. These coefficients are assumed to be constant with latitude. See text for a detailed description on how the meridional diffusion coefficients were defined. The vertical diffusion coefficient $K_{zz}$ is represented in dotted black line. The vertical CH$_{4}$ molecular diffusion coefficient is represented in black solid line.}
\label{fig:Kyy}
\end{figure}

Because of the very few constraints regarding the latitudinal and vertical dependency of this coefficient, our $K_{yy}$ have no latitudinal variations and we have scaled them over $K_{zz}$ in some cases, like for $K_{yy}^{(1)}$ and $K_{yy}^{(2)}$. We have taken the coefficient $K_{yy}^{\mathrm{(Liang)}}$ following \citet{Liang2005}. We have based $K_{yy}^{\mathrm{(Lellouch 2)}}$ on constraints derived at a pressure level of 0.3\,mbar by \citet{Lellouch2002}, \citet{Moreno2003}, \citet{Griffith2004}, and \citet{Lellouch2006}. Finally, $K_{yy}^{\mathrm{(Lellouch 1)}}$ is similar to $K_{yy}^{\mathrm{(Lellouch 2)}}$, and accounts for an additional scaling over $K_{zz}$. These meridional diffusion coefficients are displayed on Fig. \ref{fig:Kyy}.

\subsubsection{Accounting for meridional eddy diffusion with the 2D seasonal photochemical model}
\label{sss:accouting_for_Kyy}

The methodology used to account for the meridional eddy diffusion is the following. First, we run the 2D-seasonal photochemical model without meridional diffusion over several orbits so that differences in composition between two consecutive orbits fall below a threshold, as in \citet{Moses2005b} and \citet{Hue2015}. The threshold used in this work is 1\% and the subsequent number of orbits needed to reach this threshold was about 80 orbits. Once the photochemical model has reached the seasonal steady state, the meridional diffusion coefficient was turned on in the model. At this point and depending on the strength of the meridional diffusion, an additional number of orbits was necessary to reach the new seasonal steady state. The stronger the meridional diffusion coefficient, the shorter the time to reach the new seasonal steady state. As an example, the time needed for the model to reach the seasonal steady state for the diffusion coefficient presented above ranges from about 5 to 40 orbits.

\begin{figure}[!h]
\centering
{\includegraphics[width=0.45\columnwidth]{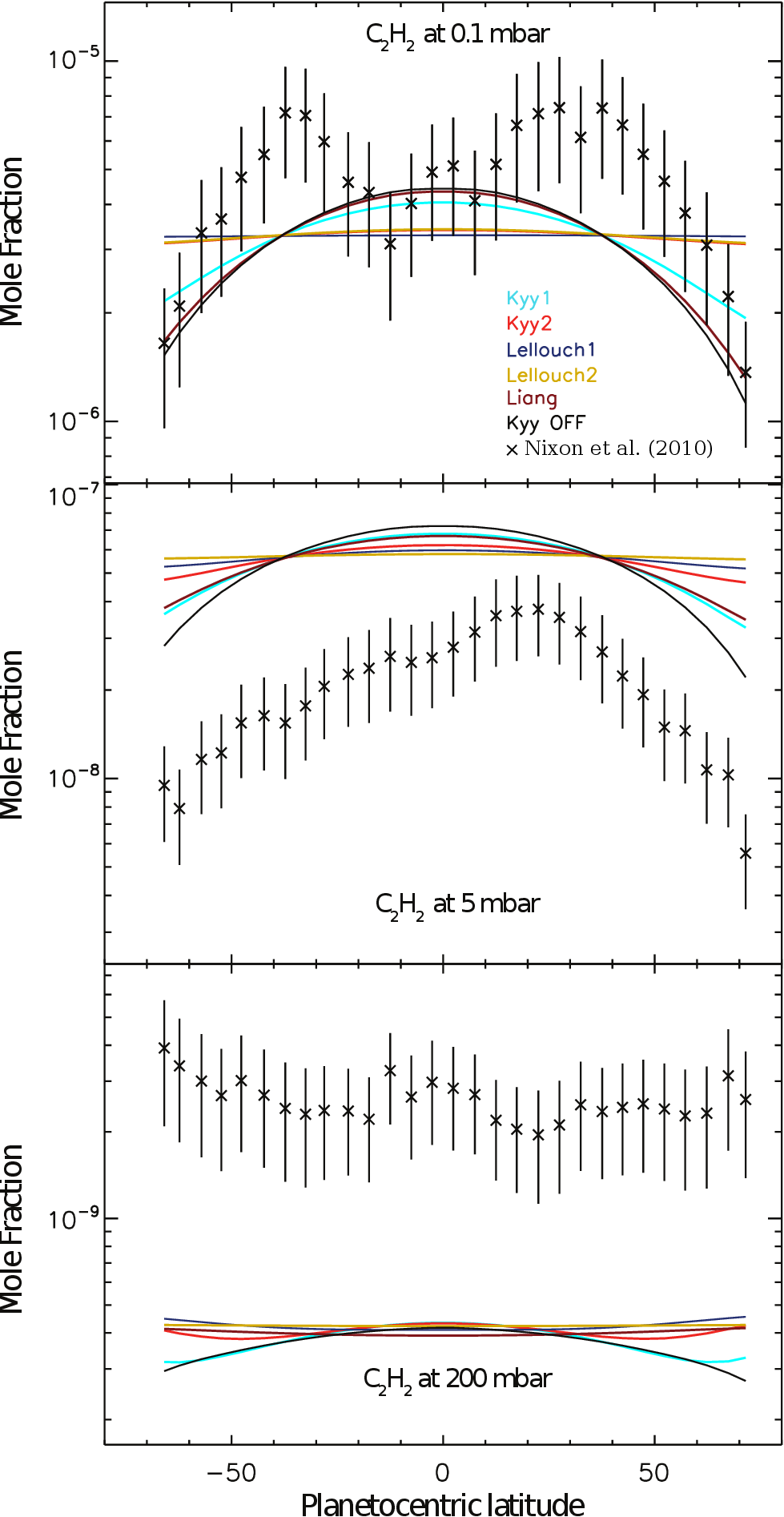}
\includegraphics[width=0.45\columnwidth]{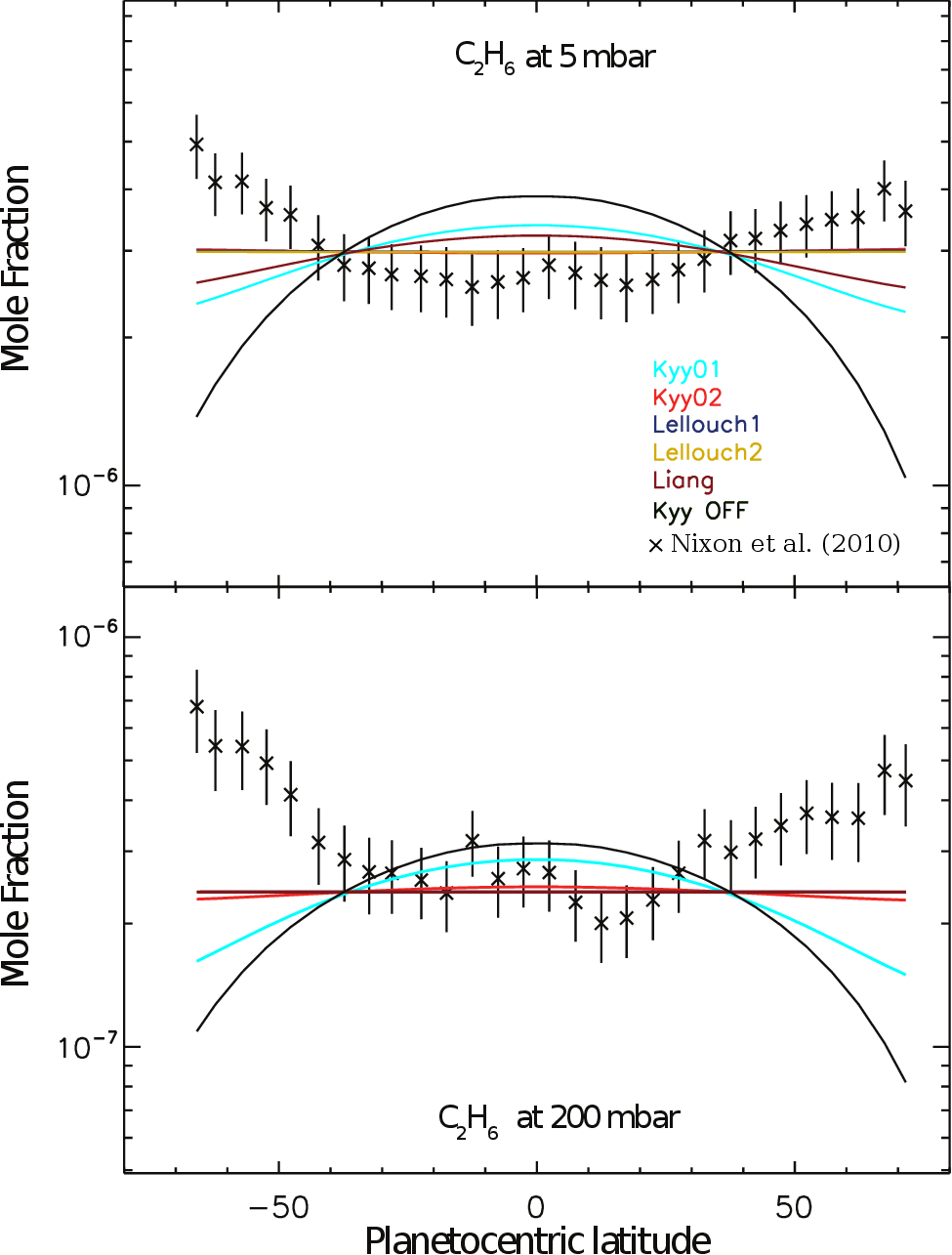}
}
\caption{Left panel: meridional distribution of C$_2$H$_2$ at 0.1\,mbar, 5\,mbar and 200\,mbar, respectively from top to bottom. Right panel: meridional distribution of C$_2$H$_6$ at 5\,mbar and 200\,mbar, from top to bottom, respectively. The 2D-seasonal photochemical model results are presented at L$_s$ = 90$^{\circ}$. Results are shown using the set of meridional diffusion coefficients presented in Fig. \ref{fig:Kyy} and Table 1 and also with no meridional diffusion coefficient. Data points are measurements from Cassini/CIRS retrieved by \citet{Nixon2010}.}
\label{fig:merid_distrib_with_Kyy}
\end{figure}

Fig. \ref{fig:merid_distrib_with_Kyy} presents the meridional distributions of C$_2$H$_2$ and C$_2$H$_6$ at the seasonal steady state at L$_s$ = 90$^{\circ}$, as obtained with the above described meridional eddy diffusion coefficients. However, due to Jupiter's very low obliquity, the choice in the L$_s$ to display the seasonal results does not affect the trends and the amplitude of the meridional distributions. The average meridional value of C$_2$H$_2$ and C$_2$H$_6$ in Fig. \ref{fig:merid_distrib_with_Kyy} do not superpose to the observations retrieved by \citet{Nixon2010}. Several effects might affect the absolute values of the hydrocarbon abundances. First, and from the observation analysis perspective, two different sets of observations might give different absolute abundances depending on the completeness of the spectral line database used to retrieve them (see for instance the section 4.2.2 of \citealt{Guerlet2009}). These absolute values are also very sensitive to systematic calibration offsets, and their error bars generally do not account for uncertainties inherited from the temperature inversions. It is also not clear how the assumed methane vertical profile used in the retrieval affects in return the averaged meridional abundances of the hydrocarbons. This might lead to additional uncertainties on the retrieved C$_2$H$_2$ and C$_2$H$_6$, given the broad range for the methane abundances around its homopause (see Fig. \ref{fig:Kzz_Jup}). In addition, and this time from the modeling perspective, the uncertainties in the photochemical models also affect the computed abundances. For instance, chemical uncertainties on the reactions producing C$_2$H$_2$ and C$_2$H$_6$ might affect their globally averaged abundances (see for instance \citealt{Dobrijevic2011}). Finally, the sensitivity of the absolute abundances to the relatively poorly constrained $K_{zz}$ is also critical in the photochemical model. We have tested several dozens of $K_{zz}$ by altering the $K_{zz}$ model C of \citet{Moses2005a} in order to better fit the average meridional value of the hydrocarbons. We have found that the $K_{zz}$ model C was the one that provided the closest agreement with respect to the observations. Now, what can be considered as robust results are the trends found in the meridional distributions, those predicted by the photochemical models, on the one hand, and those that have been retrieved from observations, on the other hand, especially if error bars are smaller than the trends themselves.

In order to quantitatively assess the relative goodness of a given $K_{yy}$, one should compute the $\chi^2$ of the model outputs. However, due to the mismatch in the absolute values between the model output and the observations, calculating the $\chi^2$ will not lead to relevant results. One could also rescale the model outputs to fit the absolute values of the observations as it is sometimes done in the literature. Rescaling should be done carefully because this assumes that most of the uncertainties both in the observations (temperature profile uncertainties, spectral database, calibration offset), and in the photochemical model (chemistry, $K_{zz}$), affect the chemical abundances linearly with latitude. To illustrate that, a change in the $K_{zz}$ will, for instance, modify the ratio between the different timescales involved in defining the meridional trends, which will therefore alter both the absolute value of the predicted abundances but also the meridional trends.
 
In this work, we have chosen not to rescale the chemical abundances but rather to compare equator-to-pole abundance ratios. Table \ref{fig:tab_merid_ratio} presents these ratios for both the observations and the photochemical models with the $K_{yy}$ presented above. Since some hemispheric asymmetry is observed in some observations, we have calculated these ratios for both hemispheres. The abundance ratios in the Northern (resp. Southern) hemisphere corresponds to the abundance ratio between latitudes of 71.5$^{\circ}$ and 2.4$^{\circ}$ (resp. -65.9$^{\circ}$ and 2.4$^{\circ}$).

\renewcommand{\arraystretch}{1.5}
\begin{table*}[t]
\footnotesize
\begin{center}
\begin{tabular}{ l l | l l l | l l l }
\hline
 & & C$_2$H$_2$ &   &   &  C$_2$H$_6$   &   &   \\
 & & 0.1\,mbar &  5\,mbar & 200\,mbar  &   5\,mbar & 200\,mbar  &   \\
\hline
\hline
Observations  & N & 0.27$\substack{+0.31 \\ -0.14}$  & 0.20$\substack{+0.20 \\ -0.10}$   &  0.91$\substack{+1.3 \\ -0.56}$         &      1.29$\substack{+0.45 \\ -0.33}$  &  1.68$\substack{+0.88 \\ -0.60 }$   &   \\
  & S &  0.32$\substack{+0.39 \\ -0.18}$ &  0.34$\substack{+0.34 \\ -0.17}$   &  1.38$\substack{+1.95 \\ -0.85}$  & 1.76$\substack{+0.61 \\ -0.45}$  & 2.54$\substack{+1.35 \\ -0.90}$   &   \\
\hline
$K_{yy}$ = 0  &N &  0.25 & 0.31  & 0.65        &     0.27      &  0.26 &   \\
  & S  & 0.35  & 0.39  &  0.70         &       0.36      &   0.35  &   \\
\hline
K$_{yy}^{(1)}$  &N &  0.48  & 0.48  &  0.76  &  0.67  &  0.54  &   \\
  & S  &  0.35   &  0.40  &  0.71          &         0.70       &     0.58    &       \\
\hline
K$_{yy}^{(2)}$  & N  & 0.91  & 0.75   & 0.96           &        1.02   &    0.91 & \\
  &  S &  0.92   & 0.76     &   0.92            &      1.02      &  0.92    &      \\
\hline
$K_{yy}^{\mathrm{(Lellouch 1)}}$  & N  &    0.99   &   0.87  & 1.11               &   1.00     & 1.00 &   \\
  &  S &   0.99  &  0.88  &  1.09  &   1.00 &  1.00   &      \\
\hline
$K_{yy}^{\mathrm{(Lellouch 2)}}$  & N  &  0.92  &  0.96  &  1.01            &          1.00    &   1.00  &  \\
  & S  &   0.92    &   0.97 &  1.01           &    1.00     &   1.00  &  \\
\hline
$K_{yy}^{\mathrm{(Liang)}}$ &N  &   0.30    &   0.52    &   1.06           &      0.78     &    1.00    &  \\
  &S    &   0.38   &   0.57    &   1.05          &    0.80     &  1.00     &    \\
\hline
\hline
\end{tabular}
\end{center}
\caption{C$_2$H$_2$ and C$_2$H$_6$ abundance ratios in the northern and southern hemispheres. These ratios represent the predicted or measured abundance values at high latitudes divided by the ones at the equator. The observed values (first row) comes from \citet{Nixon2010} and the predicted values (second row until this end) from the photochemical model developed in this work, which uses the different meridional diffusion coefficient shown in Fig. \ref{fig:Kyy}.}
\label{fig:tab_merid_ratio}
\normalsize
\end{table*}
\renewcommand{\arraystretch}{3.0}
\normalsize

\subsubsection{Results on C$_2$H$_2$}
\label{sss:C2H2_Diff}

The C$_2$H$_2$ observed abundance ratio at 0.1\,mbar and 5\,mbar is better reproduced by the model with $K_{yy}$ = 0, although the agreement is also fairly reasonable with $K_{yy}^{(1)}$ and $K_{yy}^{\mathrm{(Liang)}}$. Despite the large error bars, this tends to contradict the previous determinations of the $K_{yy}$ using post-SL9 observations ($K_{yy}^{\mathrm{(Lellouch 1)}}$ and $K_{yy}^{\mathrm{(Lellouch 2)}}$), which state that $K_{yy}$ should range between 2 and 5 $\times$ 10$^{11}$\,cm$^{2}$s$^{-1}$ around 0.2-0.5\,mbar \citep{Lellouch2002,Lellouch2006, Moreno2003,Griffith2004}. We will discuss this disagreement in section \ref{s:Discussion}. Note that the photochemical model is also not able to reproduce the local maximum of C$_2$H$_2$ abundance seen at latitudes of $\pm$ 30-40$^{\circ}$ with meridional diffusion only. 

At 200\,mbar, the observational error bars are large so that it is difficult to reach any meaningful conclusions. Indeed, all the predicted abundance ratios, from the $K_{yy}$ = 0 case to the strongest $K_{yy}$ used, are within the observational uncertainties. From a modeling point of view, it is worth noting that, when adding the meridional diffusion, several models predict a C$_2$H$_2$ abundance that increases with latitude. This is caused by the fact that C$_2$H$_2$ self-shields in the wavelength region where the methane absorption starts to decrease while the C$_2$H$_2$ absorption is still very high, i.e. around 150\,nm. When the meridional diffusion is set, C$_2$H$_2$ abundance is more abundant at high latitudes than in the case where $K_{yy}$ = 0. This stops the solar radiation in the wavelength region where C$_2$H$_2$ is photolyzed, and therefore shields C$_2$H$_2$ located at higher pressure levels (see Fig. \ref{fig:Prod_Perte_C2H2_jup}). We can also note that the increase in the abundance with respect to latitude is not observed for high value of $K_{yy}$ in the lower stratosphere (e.g. the case with $K_{yy}^{\mathrm{(Lellouch 2)}}$).

\subsubsection{Results on C$_2$H$_6$}
\label{sss:C2H6_Diff}

Our photochemical model without meridional diffusion predicts a decrease in the C$_2$H$_6$ abundance with latitude, in contradiction with the observations. Adding the meridional diffusion reduces the meridional abundance ratio and therefore increases the agreement with the observations. However, adding this cannot produce the positive meridional gradient, i.e. increasing abundances with latitude.

\subsubsection{Conclusion regarding the addition of meridional eddy mixing}
\label{sss:Discuss_Diff}

The meridional trends observed by Cassini cannot be reproduced by the 2D seasonal photochemical model with meridional eddy diffusion only. The C$_2$H$_2$ observations are fairly well reproduced even without meridional diffusion ($K_{yy}$$=$0). On the other hand, adding such diffusion improves the fit of the C$_2$H$_6$ observations. The predicted meridional trend increasingly flattens with increasing $K_{yy}$, but meridional diffusion only cannot produce an increasing abundance with latitude. An additional process is therefore needed to understand the observed meridional trends. We will now add 2D advective transport to the model.

\subsection{2D advective transport}
\label{ss:Transport}

\subsubsection{Introduction}
\label{sss:intro_transport}

Cassini/CIRS has provided observations that help to constrain the Jovian stratospheric advective transport. \citet{Lellouch2006} have retrieved the HCN and CO$_2$ meridional distributions. While the HCN column density distribution peaked at 45$^{\circ}$S, slowly decreasing northward and sharply decreasing southward, the CO$_2$ distribution, on the other hand, peaked at the south pole. Such differences are hard to explain and \citet{Lellouch2006} have reviewed the processes that might affect the production and loss rates of HCN and CO$_2$. Their conclusion is that both species resided at different altitudes, 0.5\,mbar for HCN and 5-10\,mbar for CO$_2$, and that the observed distributions were caused by a combination of meridional diffusion and advective transport that differed at these two altitudes.

In what follows, we explore how 2D advective transport can affect the meridional distributions of hydrocarbons. We should remind here that, due to the nature of the present model, 
a 2D advective transport pattern is an input to the model and not a by-product. This part should therefore be seen as an exploratory work, mainly for two reasons. First, the predictions made by dynamical models differ depending on the assumed parameters. Then, the few studies that attempted to constrain advective transport (e.g. \citealt{Lellouch2006}) could not fully define the characteristics of the advective cells. Many unknown parameters remain to fully constrain them, like the maximum amplitude of meridional winds, or the meridional and vertical extent of the cells. Here, we first expose how the advective transport pattern has been included in the photochemical model. Then, we present how advection affects the abundances.

\subsubsection{Implementation of 2D advective transport}
\label{sss:Transport_implementation}

We present here the way $v_i^{\theta}$ and $v_i^r$ have been defined in equations \ref{eq:flux_z} and \ref{eq:flux_theta}. In order to fulfill the mass-balance requirement, one has to make sure that $\mathrm{div} (n\,\textbf{v}) = 0$, where $n$ is the density and $v$ the velocity. This can be done either by deriving $n\,\textbf{v}$ from a stream function, noted $\Psi$, or directly by using GCM outputs\footnote{The Eulerian-mean-circulation obtained from GCM simulations can indeed be converted into the residual-mean-circulation, which can be used as a first approximation of the generalized Lagrangian-mean circulation (see section 3 in \citealt{Butchart2014}).}. The stream function characterizes the flow, and the flux can be calculated as $n\,\textbf{v} = \mathbf{rot( \Psi )}$. For convenience, we then define what we call a form function $\bf{B}$ defined as $\mathbf{\Psi}\,=\,n\,\textbf{B}$. The form function was defined using one or several sinusoidal functions, depending on the desired form for the advective transport.

To model a two-cell advective transport pattern, we have defined two sine functions, one for each hemisphere. Once the form and stream functions defined, the vertical and meridional winds are computed through $\textbf{v} = \mathbf{rot(\Psi)}/n$. \textbf{In addition to that, we assumed that the form function is such that the calculated wind at the boundaries of the model are null, both in the latitudinal and vertical dimension.}

\begin{figure}[!h]
\centering
{\includegraphics[width=1.0\columnwidth]{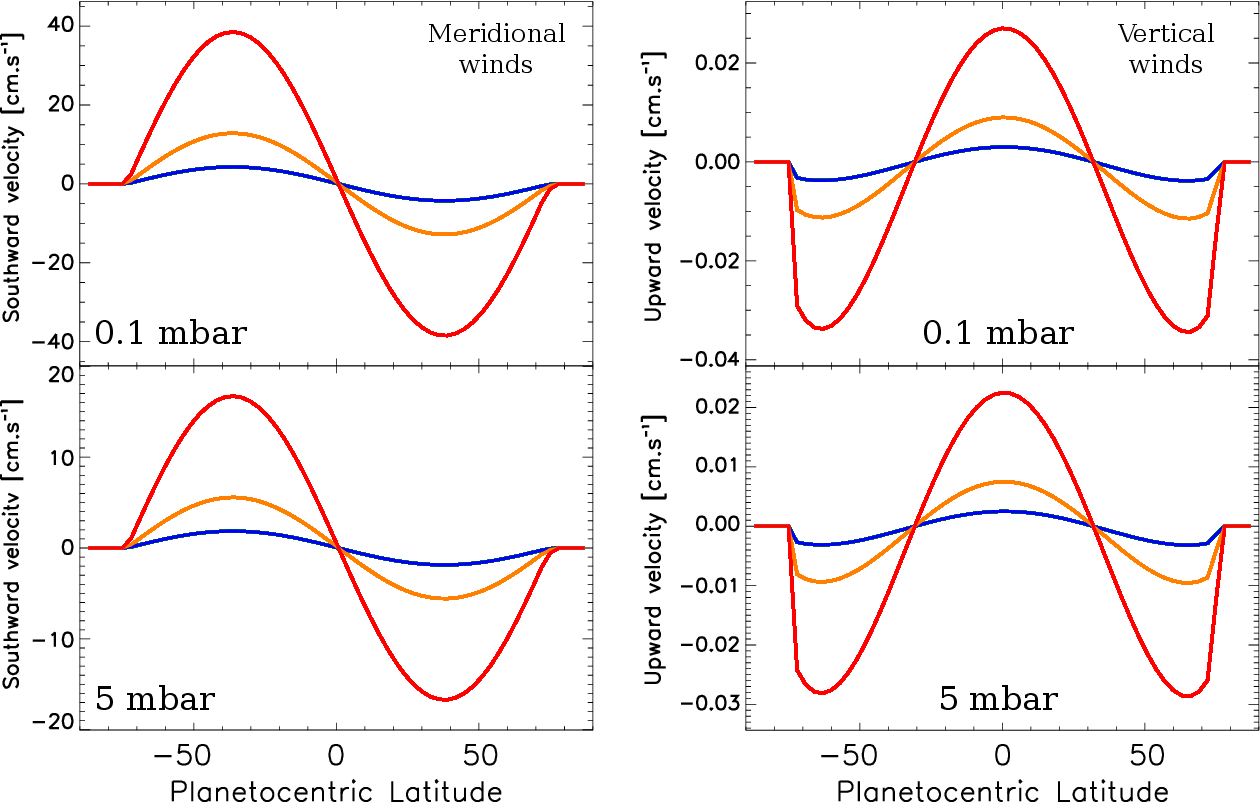}
}
\caption{Example of stratospheric 2D advection patterns added to the model. Left panel: meridional winds. Positive meridional winds are southward. Right panel: vertical winds. Positive vertical winds are upward. the upper and lower plots represent the winds at 0.1\,and 0.5\,mbar (respectively), both as a function of latitude. The different colours represent winds with various amplitude.}
\label{fig:Wind}
\end{figure}

In the framework of this exploratory section, we then rescale the obtained winds in order to test various intensities. Fig. \ref{fig:Wind} presents the meridional and vertical advection winds we have used in this work. In this example, the advective transport pattern consists of a two-cell structure, with upwelling winds at the equator and downwelling winds at higher latitudes, in both hemispheres. The azimuthal component of the stream function used to produce this advective wind field is:

\begin{align}
\Psi^{\phi} (p, \theta) =&  n F \sin \left( 2\pi\, \dfrac{\theta - \theta_{min} }{ \theta_{max} - \theta_{min} } \right) \sin \left( \pi \dfrac{ \log( p/p_{max}) }{ \log( p_{max}/p_{min})} \right)
\end{align}

where $F$ is an arbitrary constant used to rescale the wind intensities, $\theta$ the colatitude. $\theta_{min}$ and $\theta_{max}$ are the colatitude boundaries beyond which the steam function becomes null. Depending on the latitude grid used, $\theta_{min}$ and $\theta_{max}$ respectively range between 5$^{\circ}$ and 15$^{\circ}$ and 165$^{\circ}$ and 175$^{\circ}$. The pressures $p_{max}$ and $p_{min}$ delimit the pressure range for which the stream function is not null. The pressure range is defined here as $p_{max}$\,=\,$30\,$mbar down to $p_{min}$ = $10^{-2}\,$mbar. The radial and latitudinal component of the stream function are null.

Once the stratospheric advective transport pattern defined, the next step is to add it to the photochemical model. From the stationary state obtained after adding meridional eddy diffusion (see section \ref{ss:Mixing}), we have run the photochemical model with the newly defined advective transport pattern over several orbits until it fulfilled the same convergence criterion used previously in this work, i.e. when the relative variations in the hydrocarbon abundances were less than 1\% between two successive orbits. Similarly to our previous findings, the number of orbits needed to fulfill this criterion varies depending on the advective winds strengths, the pressure level, the latitude and the molecule considered. As an example, about 7 orbits were needed for the C$_2$H$_6$ abundance to converge at 200\,mbar. One should remember that, when advective transport is added to the model, photochemistry, vertical and meridional eddy diffusions are still at work.

The sensitivity of the results against the numerical diffusion has been tested by running simulations at a higher spatial resolution, in both dimensions. The nominal model uses 126 vertical levels and 35 meridional levels. Identical conclusions were drawn when the model was run at the following resolution: (i) 126 vertical levels and 59 meridional levels, (ii) 167 vertical levels and 59 meridional levels.

\subsubsection{Results with advective transport and $K_{yy}^{\mathrm{(Lellouch 2)}}$}
\label{sss:Results_transport1}

We present here results after adding 2D advective transport to the photochemical model previously run with $K_{yy}^{\mathrm{(Lellouch 2)}}$. This $K_{yy}$ was preferred over the other ones because (i) there are multiple observational evidences pointing toward this value of $K_{yy}$ (see section \ref{ss:Mixing}) and (ii) over all the $K_{yy}$ used in our study, this one provides the closest fit to the C$_2$H$_6$ Cassini observations. 

\begin{figure}[!h]
\centering
{\includegraphics[width=0.45\columnwidth]{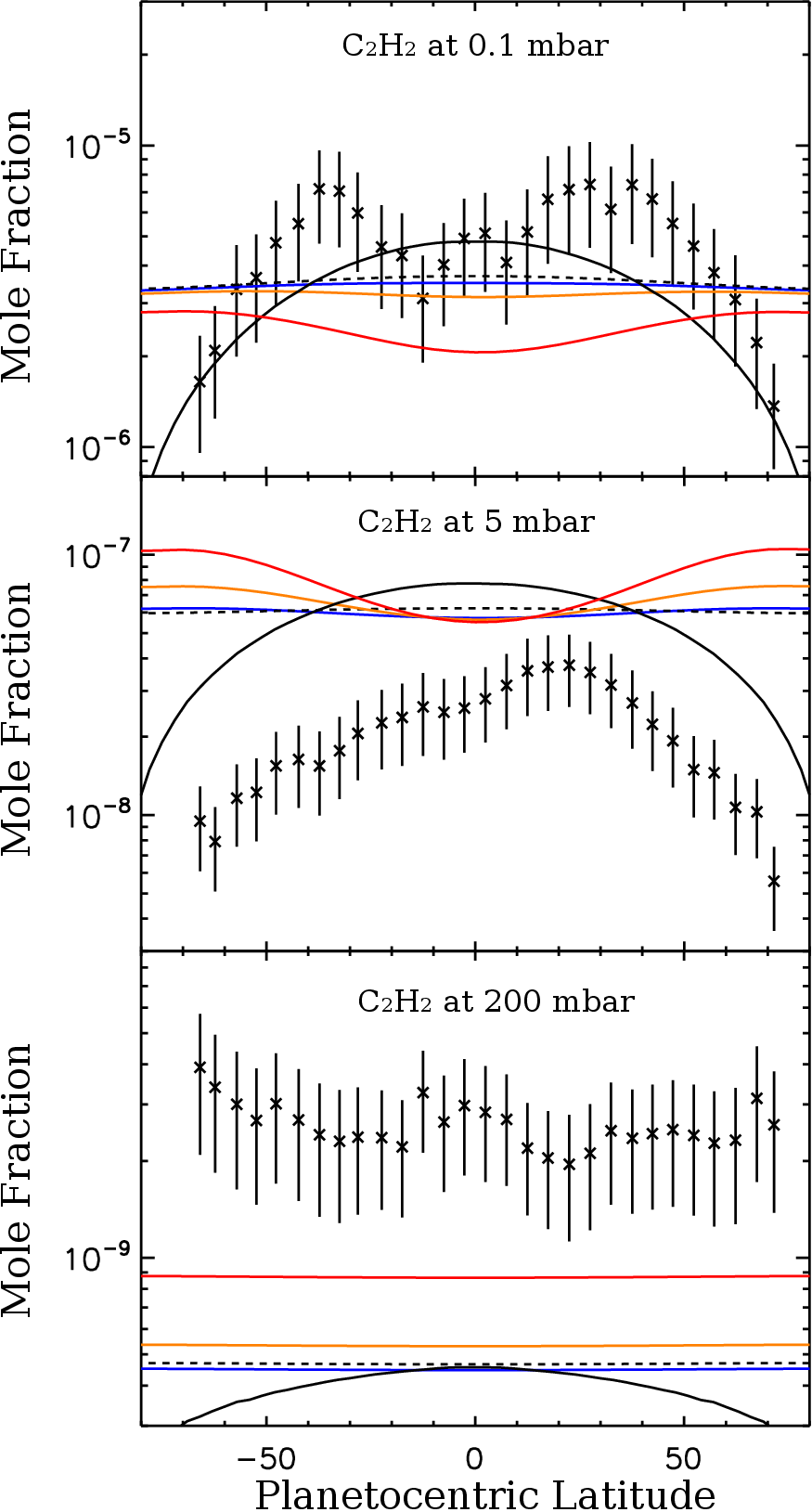}
\includegraphics[width=0.45\columnwidth]{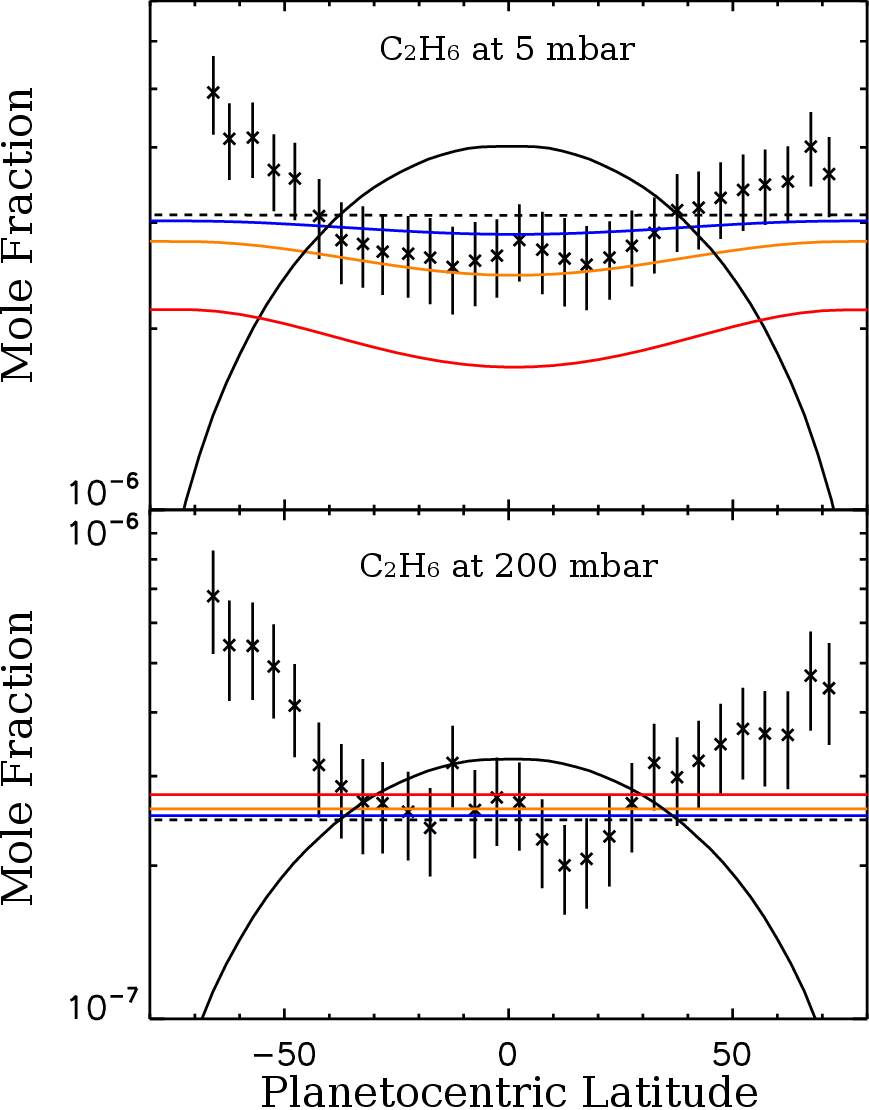}
}
\caption{Meridional distribution of C$_2$H$_2$ and C$_2$H$_6$ abundances at 0.1\,mbar (upper plot), 5\,mbar (middle plots) and 200\,mbar (lower plots). Black solid line: photochemical predictions with $K_{yy}$ = 0. Black dashed line: photochemical predictions with $K_{yy}$ = $K_{yy}^{\mathrm{(Lellouch 2)}}$ (see section \ref{ss:Mixing}). Colored lines: photochemical predictions with $K_{yy}$ = $K_{yy}^{\mathrm{(Lellouch 2)}}$ and stratospheric 2D advective transport. The various wind fields used are presented in Fig. \ref{fig:Wind}. The model results are compared with the Cassini/CIRS observations of Jupiter \citep{Nixon2010}.}
\label{fig:C2H2_C2H6_cut_wind}
\end{figure}

The effect of the stratospheric 2D advection transport, as defined in Fig. \ref{fig:Wind}, on the meridional distributions of C$_2$H$_2$ and C$_2$H$_6$ are presented on Fig. \ref{fig:C2H2_C2H6_cut_wind}. The main effect on C$_2$H$_2$ and C$_2$H$_6$ resulting from the addition of the two-cells pattern with upwelling winds at the equator and downwelling winds at higher latitude is to invert the meridional trends. Indeed, the abundances now increase from the equator towards higher latitudes. This inversion depends on the intensity of the winds: the stronger the winds, the stronger the inversion. The wind cells extend from 30\,mbar up to 10$^{-2}$\,mbar. Therefore, the effects of the advective transport on the 200\,mbar abundance are only caused by the vertical and meridional eddy diffusion.

\begin{figure}[!h]
\centering
{\includegraphics[width=1.0\columnwidth]{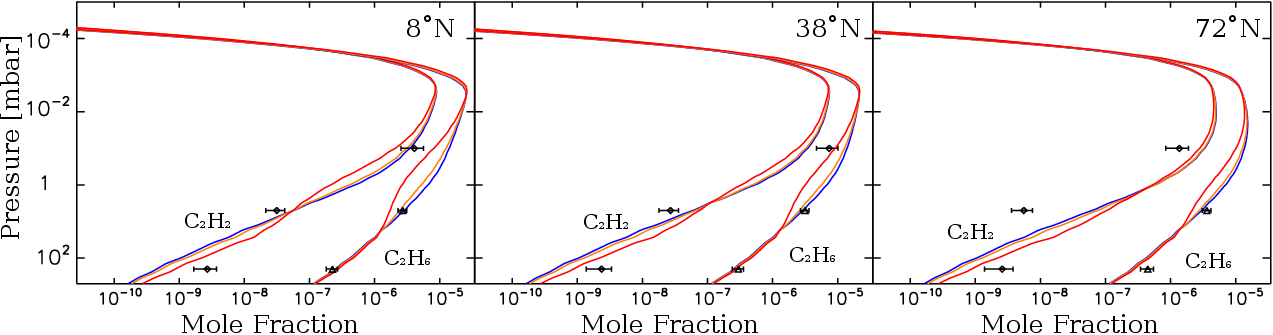}
}
\caption{Vertical profiles of C$_2$H$_2$ and C$_2$H$_6$ at 8$^{\circ}$N (left panel), 38$^{\circ}$N (middle panel) and  38$^{\circ}$N (right panel). The vertical profiles are only presented when the stratospheric circulation is added to the photochemical model. The corresponding wind velocities are presented on Fig. \ref{fig:Wind}. The model results are compared to the Cassini/CIRS observations of Jupiter at 0.1\,mbar (C$_2$H$_2$ only), 5\,mbar and 200\,mbar \citep{Nixon2010}.}
\label{fig:C2H2_C2H6_wind_vertical}
\end{figure}

Fig. \ref{fig:C2H2_C2H6_wind_vertical} presents the vertical profiles of C$_2$H$_2$ and C$_2$H$_6$ when adding the effects of advective transport. They are presented at three distinctive latitudes observed with Cassini/CIRS. 

For pressures ranging between 10$^{-2}$\,mbar and 5-10\,mbar, the upwelling winds in the equatorial zone cause a decrease in these two hydrocarbon abundances when compared to a case without meridional circulation. Indeed, these upwelling winds bring air from higher pressure level where the hydrocarbons are less abundant. At higher latitudes, one would expect the opposite behavior to occur, i.e. due to the downwelling winds which bring hydrocarbon enriched air from higher altitudes, the hydrocarbon abundances should be increased. However, due to the coupling with the meridional diffusion coefficient, which tends to remove compositional gradients, the atmosphere is still impoverished in hydrocarbon at these latitudes. Note, from Figs. \ref{fig:C2H2_C2H6_cut_wind} and \ref{fig:C2H2_C2H6_wind_vertical}, that the decrease of the hydrocarbon abundances is less pronounced at high latitudes than in the equatorial region.

In the lower stratosphere, the opposite situation is observed. The downwelling winds at high latitude bring air enriched in hydrocarbons from higher altitudes. Due to the coupling with the meridional diffusion coefficient, the hydrocarbon over-abundance, with respect to a case without winds, diffuses toward the equator, also causing an over-abundance in this region. This over-abundance then diffuses to even higher pressure levels, where no winds were defined. 

Adding the stratospheric 2D advective transport affects the hydrocarbon abundances, even in atmospheric regions where the winds were set to zero. This is especially true for the lower stratosphere because these regions are coupled to the upper stratosphere through the vertical diffusion. 

It is very interesting to note that, due to the difference in the photochemical lifetimes of C$_2$H$_2$ and C$_2$H$_6$ and due to the different vertical gradients these two species have, they react slightly differently to the coupled diffusion-advection stratospheric transport. In the upper stratosphere, between 1\,mbar and 10$^{-2}$\,mbar, C$_2$H$_6$ seems to be more affected by advective transport than C$_2$H$_2$. This is counter-intuitive because C$_2$H$_2$ has a steeper vertical gradient than C$_2$H$_6$, and should therefore be more affected by advective transport. However, one should remember that these species still undergo chemical reactions and that C$_2$H$_2$ has a shorter chemical lifetime in this region than C$_2$H$_6$. On the other hand, in the lower stratosphere, from 1\,mbar to about 100\,mbar, C$_2$H$_2$ is more affected than C$_2$H$_6$ by advective transport, consistent with its steeper vertical gradient.

\subsubsection{Sensitivity to $K_{yy}$}
\label{sss:Sensitivity_Kyy}

\begin{figure}[!h]
\centering
{\includegraphics[width=0.45\columnwidth]{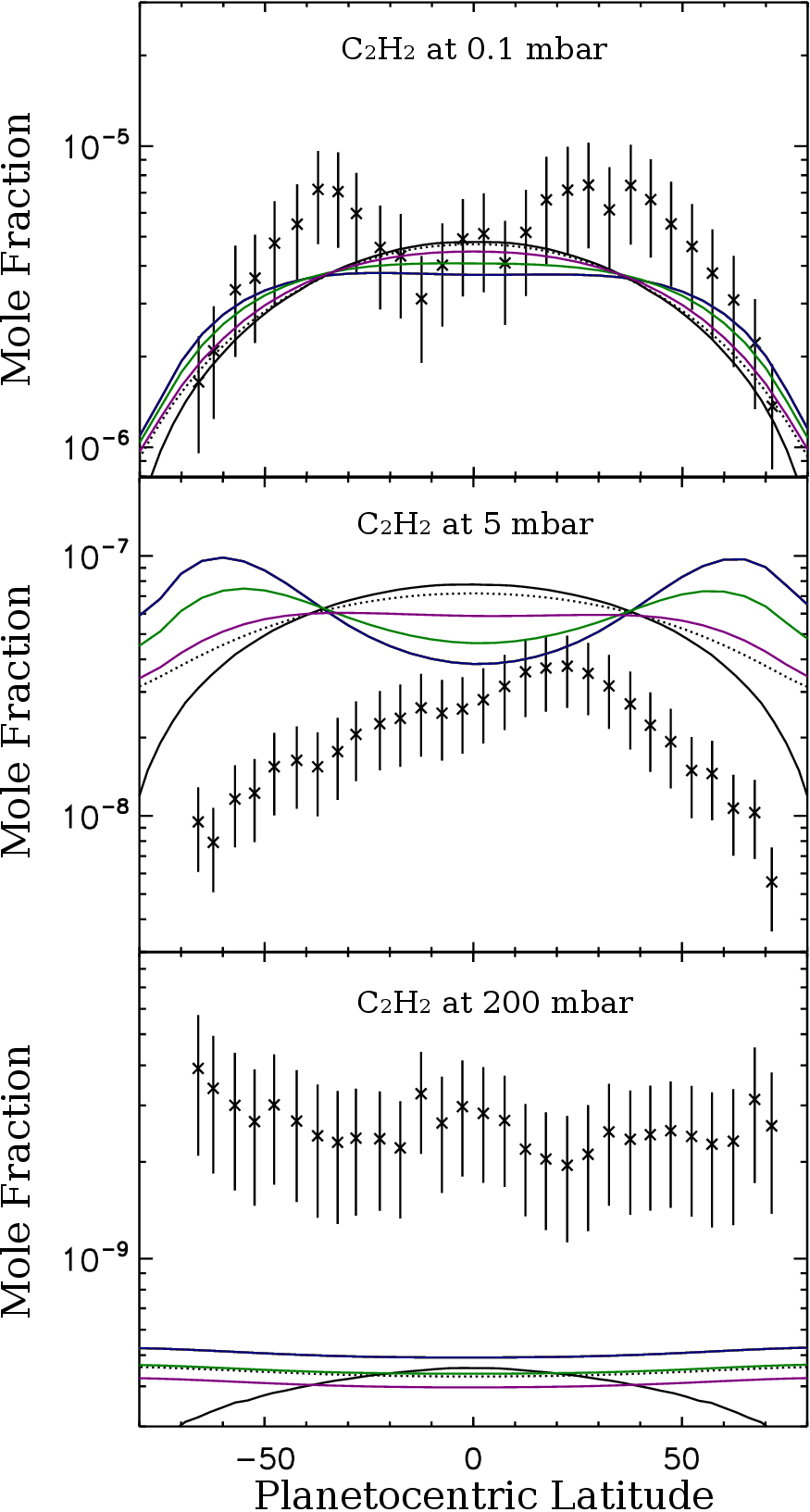}
\includegraphics[width=0.45\columnwidth]{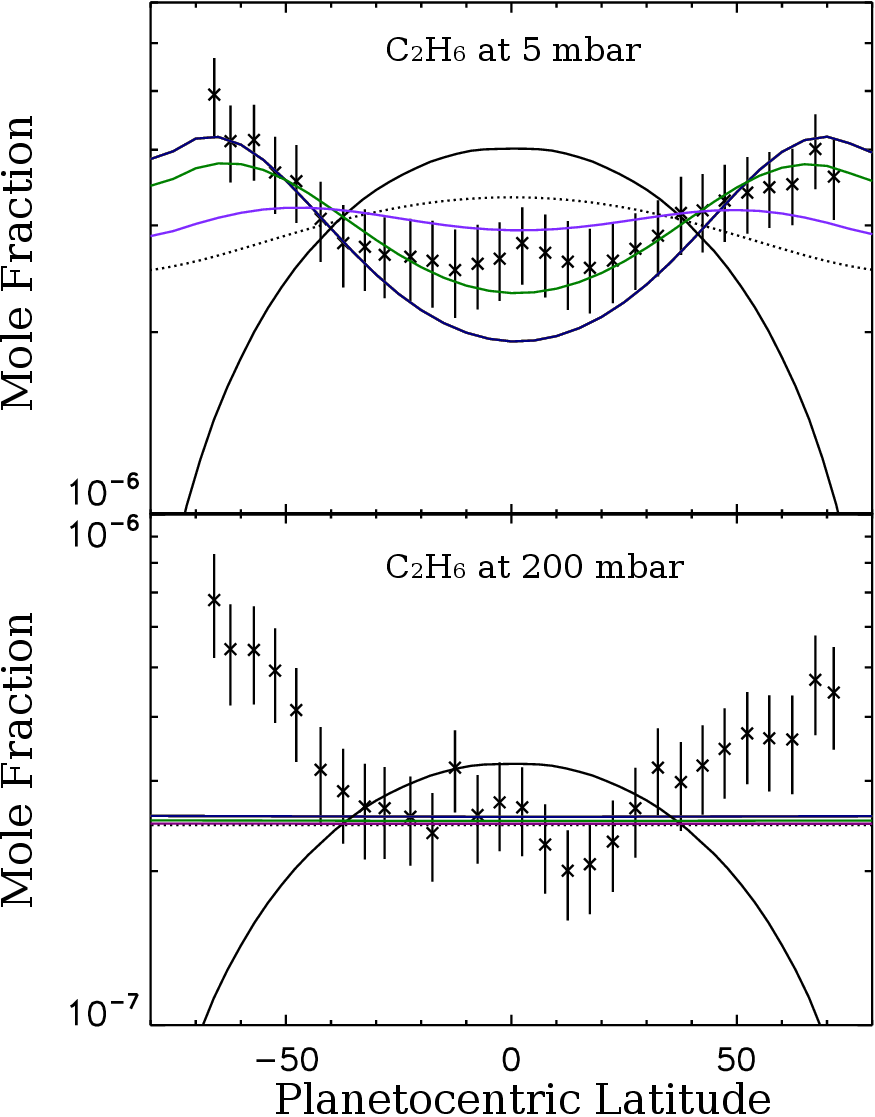}
}
\caption{Same as Fig. \ref{fig:C2H2_C2H6_cut_wind} except that (i) the wind field used is the one presented on Fig. \ref{fig:Wind_Liang}, (ii) the meridional diffusion coefficient used is $K_{yy}^{\mathrm{(Liang)}}$ (see section \ref{ss:Mixing}). Black dotted and solid lines: photochemical predictions with $K_{yy}$ = $K_{yy}^{\mathrm{(Liang)}}$ and $K_{yy}$ = 0, respectively, and without stratospheric advective transport. Colored lines: photochemical predictions with $K_{yy}$ = $K_{yy}^{\mathrm{(Liang)}}$ and stratospheric advective transport.}
\label{fig:C2H2_C2H6_cut_wind_Liang}
\end{figure}

\begin{figure}[!h]
\centering
{\includegraphics[width=1.0\columnwidth]{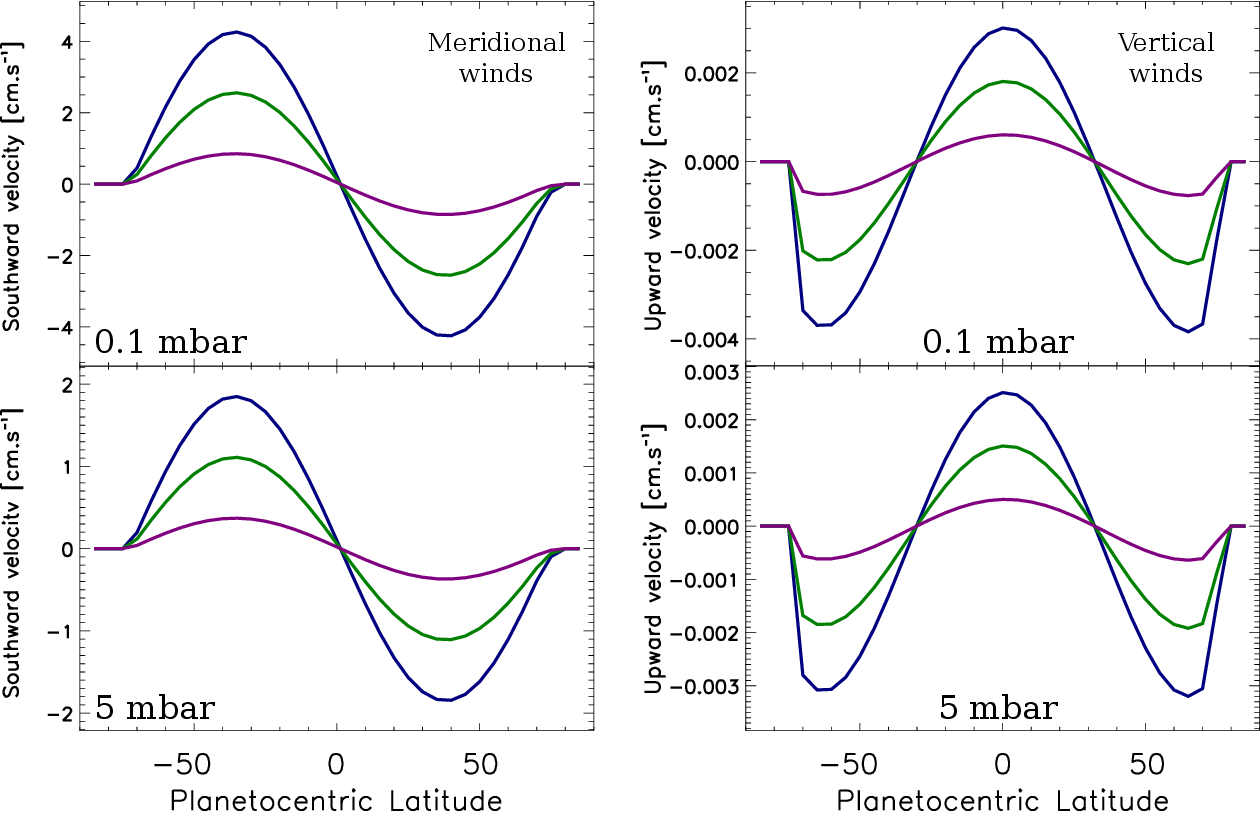}
}
\caption{Corresponding wind field used to obtain the results presented on Fig. \ref{fig:C2H2_C2H6_cut_wind_Liang}. The blue solid lines correspond to the case of the weak wind field presented on Fig. \ref{fig:Wind}.}
\label{fig:Wind_Liang}
\end{figure}

One of the most critical point in this section is the very high sensitivity of the results to the assumed meridional eddy diffusion. Using the meridional diffusion $K_{yy}^{\mathrm{(Liang)}}$ instead of $K_{yy}^{\mathrm{(Lellouch 2)}}$, the amplitude of the advective winds needed to reproduce the C$_2$H$_6$ meridional distribution at 5\,mbar is greatly reduced. Fig. \ref{fig:C2H2_C2H6_cut_wind_Liang} presents the meridional distribution of C$_2$H$_2$ and C$_2$H$_6$ abundances with the $K_{yy}^{\mathrm{(Liang)}}$ and using advective winds similar in shape, though weaker in amplitude, to the one presented in Fig. \ref{fig:Wind} (see Fig. \ref{fig:Wind_Liang}). In this case, the field with the strongest winds (in blue lines) corresponds to the weakest winds of Fig \ref{fig:Wind}. For the same wind strength as in the calculations that used $K_{yy}^{\mathrm{(Lellouch 2)}}$ (blue solid curves on Figs. \ref{fig:C2H2_C2H6_cut_wind} and \ref{fig:C2H2_C2H6_cut_wind_Liang}), the produced meridional gradients are increasingly enhanced with decreasing $K_{yy}$. Indeed, $K_{yy}^{\mathrm{(Liang)}}$ is smaller than $K_{yy}^{\mathrm{(Lellouch 2)}}$ at all pressure levels, which decreases the meridional homogenisation caused by diffusion and therefore allows the hydrocarbon meridional trends to be shaped more easily by advective transport. As an example, the maximum strength of the vertical winds needed to reproduce the C$_2$H$_6$ meridional distribution observed at 5\,mbar ranges from 35\,mm\,s$^{-1}$ when considering $K_{yy}^{\mathrm{(Lellouch 2)}}$, to about 2\,mm\,s$^{-1}$ when using $K_{yy}^{\mathrm{(Liang)}}$. On the other hand, as previously noted, this causes a stronger disagreement with the C$_2$H$_2$ observations. It is worth noting that, at 0.1\,mbar, the chemistry-diffusion-advection coupling is such that the meridional distributions are not as affected by the diffusion and advection as when using the wind field presented in Fig. \ref{fig:Wind_Liang} in combination with the $K_{yy}^{\mathrm{(Lellouch 2)}}$. This suggests that, at 0.1\,mbar, the meridional diffusion coefficient proposed by \citet{Lellouch2006} either overestimates the strength of the meridional diffusion, or that an unknown process is coupled to the C$_2$H$_2$ chemistry.

\subsubsection{Caveats}
\label{sss:caveats}

The main purpose of this section is to demonstrate that the observed distribution of C$_2$H$_6$ at 5\,mbar can be explained when accounting for 2D advective transport, with upwelling winds at the equator and downwelling winds at higher latitudes. We have shown that the amplitude of these winds depends on the strength of the associated $K_{yy}$. 
However, several caveats have to be carefully kept in mind. First, and probably the most important one, the fact that both C$_2$H$_6$ and C$_2$H$_2$ keep following similar meridional trends, when meridional diffusion and 2D advective transport are included in the model, suggests that another mechanism (at least) that decouples the behavior of these two species is at work. This will be discussed in section \ref{s:Discussion}. Then, the extreme sensitivity of the meridional distributions to both meridional diffusion and 2D advective transport has to be reminded. These two model parameters remain poorly known. Although post-SL9 observations have resulted in some constraints on the meridional eddy diffusion coefficient, its vertical and meridional variability is not known. 

Another information worth having in mind when retrieving the meridional diffusion coefficient from observations with low spatial resolution with respect to Jupiter's size is that, 2D advective transport, which, ultimately, mixes the atmosphere, could be mistaken for diffusive processes when considered over a large (i.e. planetary) scale. Therefore, on some low spatial resolution observations, some of the information regarding advective transport could be accounted for in meridional and/or vertical diffusive processes. It is however interesting to note that the meridional diffusion coefficient retrieved from low spatial resolution observations in the millimeter range \citep{Moreno2003} was consistent with higher spatial resolution observations in the infrared range \citep{Lellouch2002, Griffith2004, Lellouch2006}.

\section{Discussion}
\label{s:Discussion}

Cassini/CIRS has revealed drastically different meridional distributions for C$_2$H$_2$ and C$_2$H$_6$. These distributions are puzzling because C$_2$H$_2$ and C$_2$H$_6$ are expected to be tightly coupled, according to our current knowledge of giant planet stratospheric photochemistry. The 2D-seasonal photochemical model of Jupiter presented here has allowed us to test several 2D diffusive and/or advective transport patterns in order to try to reconcile these observations. However, adding 2D transport does not result in a better agreement with observations. Although these two species have different evolution timescales and vertical distributions, their modelled meridional distributions remain coupled and evolve in a similar way when 2D advective transport and meridional diffusion are added to the model. These results seem in disagreement with the work of \citet{Zhang2013a} which demonstrates that a coupled chemical-diffusion-advection model would be able to reproduce opposite meridional distributions for species that have very different chemical evolution timescales, if the transport timescale is correctly chosen. This disagreement most likely comes from the fact that they used an idealized chemical network through parametrized chemical loss rates which cannot reproduce a non-linearly coupled chemical network. Our work rather suggests that 2D diffusive and advective transport alone seems unlikely to produce opposite meridional distributions for C$_2$H$_2$ and C$_2$H$_6$.

At this point, a way to produce radically different meridional distributions for these two compounds would be to invoke a (photo-)chemical process that would be strongly coupled to only one of them. Ion-neutral chemistry could be a good candidate. The vast majority of the giant planet photochemical models neglect ion-neutral chemistry, as their main purpose is to properly reproduce the globally averaged or low-spatial resolution observations of neutral species (e.g. \citealt{Ollivier2000, Moses2000a, Moses2000b, Dobrijevic2011}). 

In the thermosphere, above the methane homopause, diffusive separation causes a drop in the methane abundance as well as the other species except the lightest ones. In this region, the only species left are H$_2$, He and H. The incoming charged particles or UV/EUV light produced either by the Sun or within the magnetosphere ionise the neutral species, triggering a set of ion-neutral chemical reactions. Once produced, these ions diffuse downwards, to the methane homopause \citep{Kim1994}. The abundance of the heavier hydrocarbons begins to increase when approaching this region, and ion-neutral chemical reactions lead to the production of heavier hydrocarbon ions. \citet{Kim1994} predict that the ion-neutral reactions and the dissociative recombinations lead to the production of a stable hydrocarbon ion layer around the methane homopause. Although they have not presented the vertical profiles from the neutral atmosphere, they draw the following conclusions:

\begin{itemize}
\item the main hydrocarbon ions produced in the ion layer are CH$_5^+$, C$_2$H$_3^+$, CH$_3^+$ and C$_2$H$_5^+$,
\item the reactivity of saturated hydrocarbons (CH$_4$ and C$_2$H$_6$) with saturated or nearly saturated hydrocarbon ions (e.g. C$_2$H$_4^+$, C$_2$H$_5^+$ and C$_2$H$_6^+$) is weak,
\item the reactivity of C$_2$H$_2$ with saturated or unsaturated hydrocarbon ions is important,
\item the ion-neutral reactions lead to the formation of increasingly complex hydrocarbons.
\end{itemize}

\noindent
One of the main conclusions of the work from \citet{Kim1994} is that ion chemistry may be an important destruction mechanism for hydrocarbons and especially C$_2$H$_2$.

Over the vast majority of predicted ion species, only H$_3^+$ has been detected so far on Jupiter \citep{Drossart1989}. Although several emission features from neutral hydrocarbons have been detected around the methane homopause (e.g. \citealt{Dols2000}), there is currently no observational constraints on the hydrocarbon ion abundances predicted by photochemical models \citep{Yelle2004}.

Recent additional analysis of the Cassini/CIRS dataset by \citet{Sinclair2016} of the Jupiter flyby suggest a somehow different but complementary story for the stratospheric C$_2$H$_2$ and C$_2$H$_6$. They retrieved the 2D (latitude-longitude) distribution of C$_2$H$_2$ and C$_2$H$_6$ abundances at 2\,mbar at the time of the Jupiter flyby. \citet{Sinclair2016} showed that C$_2$H$_2$ was enhanced around Jupiter's main auroral region, at 70$^{\circ}$N (i.e. the main oval, roughly centered around 180$^{\circ}$W). On the other hand, they found that C$_2$H$_6$ was enhanced at all longitudes away from the main oval at this latitude. The retrieval method differs from the one of \citet{Nixon2007}, in a way that Nixon et al. have retrieved zonally averaged maps of temperature and hydrocarbon abundances, excluding the auroral region (60$^{\circ}$-70$^{\circ}$S, 330$^{\circ}$-90$^{\circ}$W and 60$^{\circ}$-70$^{\circ}$N, 150$^{\circ}$-210$^{\circ}$W), which therefore explains the apparent inconsistency with \citet{Sinclair2016}. Moreover, recent ground-based observations using IRTF/TEXES from \citet{Fletcher2016} confirmed the low- to mid-latitude (up to $\pm$ 60$^{\circ}$) behavior of C$_2$H$_2$ and C$_2$H$_6$.

This observed decoupling between the C$_2$H$_2$ and C$_2$H$_6$ is not understood given our current knowledge of the photochemistry in giant planets. Analogies with Titan can be used to evaluate, to a certain extent, the impact of the ion-neutral chemistry on another neutral atmosphere. Indeed, numerous ion-neutral photochemical models of Titan have been developed in order to understand the Cassini results (e.g. \citealt{Banaszkiewicz2000}, \citealt{Wilson2004}, \citealt{DeLaHaye2008a}, \citealt{Krasnopolsky2009}, \citealt{Dobrijevic2016}, and also \citet{Mandt2012} for a more complete list of Titan's photochemical models). The photochemical model of \citet{DeLaHaye2008a}, for example, suggests that C$_2$H$_2$ and C$_2$H$_4$ are more likely produced in a high electron precipitation environment than C$_2$H$_6$ (see discussion in \citealt{Sinclair2016}). Indeed, the production rates of the ion-neutral and electron recombination paths of the latter compound are several orders of magnitude lower than those of the former compounds. However, on Jupiter's auroral regions, the electron environment is expected to be different from the one on Titan. In addition to that, N$_2$ will not be competing with CH$_4$ to use the incoming electrons and produce light ions, as in Titan, given that these two species have roughly the same ionization cross-section in the 100$\,$eV range \citep{DeLaHaye2008b}. This analogy should be taken carefully, and to answer the complex question of the impact of ion-neutral chemistry on the neutral composition, the next step of this work is to implement a coupled ion-neutral chemistry in the photochemical model.

\section{Conclusion}
\label{s:Conclusion}

The Cassini/CIRS data taken during the flyby of Jupiter have revealed strikingly different meridional distribution of C$_2$H$_2$ and C$_2$H$_6$. Until now, these hydrocarbons were thought to be tightly coupled by (photo)-chemical reactions, so that the observed meridional distributions remain puzzling. Previous studies have raised the possibility that diffusive and advective transport coupled to chemistry might explain the observed distributions \citep{Liang2005, Lellouch2006}. To investigate these effects, we have developed the first 2D-seasonal photochemical model of Jupiter that accounts for 2D diffusive and advective transport. To decrease the computational time of such a model, we have used a reduced chemical network that allows us to compute the abundances of the light hydrocarbons only, while staying within the model uncertainties (see section \ref{sss:Chemical_network}). We summarize here the main results from our study.

When the meridional diffusion and 2D advective transport circulation are firstly not included, the seasonal variations of stratospheric composition are mostly controlled by Jupiter's eccentric orbit. Contrary to Saturn \citep{Moses2005b,Hue2015}, Jupiter's low obliquity plays a minor role on the seasonal modulation of its chemical composition. In this case, only the predicted meridional distribution of C$_2$H$_2$ is in fairly good agreement with the Cassini observations published by \citet{Nixon2010}. 

Several studies have estimated the strength of the meridional diffusion in Jupiter's stratosphere. These studies were primarily based on the post-SL9 observations of species deposited by the comet, that spread over Jupiter's stratosphere \citep{Lellouch2002, Moreno2003, Griffith2004}, as well as recent Cassini/CIRS observations of hydrocarbons \citep{Kunde2004, Liang2005, Lellouch2006}. When using a meridional eddy diffusion coefficient derived from observations of post-SL9 species, C$_2$H$_6$ is better reproduced than in the previous case (i.e without meridional diffusion), but the fit to the observed C$_2$H$_2$ distribution worsens. Moreover, adding meridional diffusion still cannot produce an increase of the C$_2$H$_6$ abundance with latitude. Meridional diffusion alone is therefore not sufficient to explain the meridional distributions of C$_2$H$_2$ and C$_2$H$_6$ in the stratosphere of Jupiter.

Finally, the addition of 2D advective transport cells, with upwelling winds at the equator and downwelling winds at higher latitudes, can produce a C$_2$H$_6$ meridional distribution that is in agreement with the Cassini observations. On the other hand, the C$_2$H$_2$ meridional distribution follows the same behavior as C$_2$H$_6$, and is therefore in disagreement with the observations. Tuning advective transport to increase the agreement with the C$_2$H$_6$ data will always increase the disagreement with the C$_2$H$_2$ observations. We stress once again that the meridional diffusion will counter-act the effects of advective transport. 

The very different meridional distributions of C$_2$H$_6$ and C$_2$H$_2$, which are thought to have a tied chemical history, remains puzzling. New observations with the James Webb Space Telescope should help us to monitor the seasonal evolution of their meridional distributions \citep{Norwood2016} and prepare for the future Jupiter Icy Moon Explorer mission. The present work has explored a new approach in order to explain these behaviors, by adding meridional eddy diffusion and 2D advective transport to a seasonal 2D photochemical model of Jupiter. The next step of this study will consist in accounting for ion-neutral chemistry to explore how such chemistry can impact differently the abundances of C$_2$H$_2$ and C$_2$H$_6$ at high latitudes.

\section{Acknowledgement}

Part of this work has been supported by the \textit{Investissements d'avenir} program from the \textit{Agence Nationale de la Recherche} under the reference ANR-10-IDEX-03-02 (IdEx Bordeaux). T. Cavali\'{e} acknowledges funding from the Programme National de Plan\'etologie of CNRS/INSU. We thank R. Gladstone, L. Fletcher, A. Spiga, and P. Romani for helpful comments, information and suggestions on this work. We thank Dr. Bruno B\'{e}zard and an anonymous reviewer for they constructive comments that help us improve our manuscript.


\end{document}